\documentclass[%
 reprint,
groupedaddress,
 amsmath,amssymb,
 aps,
prb,
]{revtex4-1}

\usepackage{graphicx}
\usepackage{dcolumn}
\usepackage{bm}

\begin{document}

\title{Cation spin and superexchange interaction in oxide materials  below and above spin crossover  under high pressure.}
\author{Vladimir A. Gavrichkov}
\email[]{gav@iph.krasn.ru}
\affiliation{Kirensky Institute of Physics, Akademgorodok 50, bld.38, Krasnoyarsk, 660036 Russia}

\author{Semyon I. Polukeev}
\affiliation{Kirensky Institute of Physics, Akademgorodok 50, bld.38, Krasnoyarsk, 660036 Russia}

\author{Sergey G. Ovchinnikov}
\affiliation{Kirensky Institute of Physics, Akademgorodok 50, bld.38, Krasnoyarsk, 660036 Russia}
\date{\today}

\begin{abstract}

We derived simple rules for the sign of $180^\circ$ superexchange interaction based on the multielectron calculations of the superexchange interaction in the transition metal oxides that are valid both below and above spin crossover under high pressure. The superexchange interaction between two cations in  d$^n$ configurations is given by a sum of partial contributions related to the electron-hole virtual excitations to the different states of  the d$^{n + 1}$  and  d$^{n - 1}$ configurations. Using these rules, we have analyzed the sign of the $180^\circ$ superexchange interaction of a number of oxides with magnetic cations in electron configurations from  d$^2$ till d$^8$: the iron, cobalt, chromium, nickel, copper and manganese oxides with increasing pressure. The most interesting result concerns the magnetic state of cobalt and nickel oxides CoO, Ni$_2$O$_3$ and also La$_2$CoO$_4$, LaNiO$_3$ isostructural to well-known high-T$_C$ and colossal magnetoresistance materials. These oxides have
a spin $\frac{1}{2}$  at the high pressure. Change of the interaction from antiferromagnetic below spin crossover to ferromagnetic above spin crossover is predicted for oxide materials with cations in  d$^5$(FeBO$_3$)  and   d$^7$(CoO) configurations, while for materials with the other d$^n$  configurations spin crossover under high pressure does not change the sign of the $180^\circ$ superexchange interaction.

\end{abstract}


\maketitle

\section{\label{sec:intr}Introduction\\}
The mechanism of superexhcange interaction is well known for a long time \cite{Anderson1959}. The effective Heisenberg Hamiltonian describes the exchange interaction  $J$ of the magnetic cations in the ground state. It is well known that there are many excited states for multielectron cations.~\cite{Tanabe1956} However these states are not involved by the superexchange interaction, and Heisenberg model is a based theory because typically excited states lies well above the magnetic scale  $J$  and Curie/Neel  temperatures (${{{T_C}} \mathord{\left/
 {\vphantom {{{T_C}} {{T_N}}}} \right.
 \kern-\nulldelimiterspace} {{T_N}}}$). A low energy description of magnetic insulators may be violated in two situations. The first one is related with intensive optical pumping when one of magnetic cations is excited into some high energy state and its exchange interaction with the neighbor cation in the ground state changes~\cite{Mentink2014} resulting in many interesting effects of the femtosecond magnetism. \cite{Kalashnikova2015, Gavrichkov2017} The other situation occurs at the high pressure when the cation spin crossover in magnetic insulators from the high spin (HS) to the low spin (LS) state takes place.~\cite{Tanabe1956, Lamonova2011} 
The spin crossover occurs due to competition between the energy of the crystalline field $10Dq$ and the parameter of intratomic Hund exchange $J_H$. Typically, the applied pressure increases the crystal field, but does not significantly change the exchange parameter $J_H$.
The spin crossovers are known for many transition metal oxides with   ${d^4} \div {d^7}$ cations, and for transition metal complexes, like metalorganic molecules or molecular assemblies.~\cite{Lyubutin2009, Halder2002,Brooker2009,Ohkoshi2011,Wentzcovitch2009,Hsu2010,Liu2014,Sinmyo2017, Marbeuf2013, Nishino2007, Ovchinnikov2004, Saha2014} Near crossover the energies of two states  ${\varepsilon _{HS}}$ and   ${\varepsilon _{LS}}$ are close to each other and conventional scheme of the superexhcange interaction calculation should be modified.
 

The spin crossovers have been experimentally detected and investigated in a number of transition metal oxides.~\cite{Lyubutin2009} Calculations also confirm a possibility of the spin crossovers in these materials and their role in the metal-insulator transition.~\cite{Gavrichkov2017, Gavrichkov2018} In real, a situation is complicated by the observed structural and chemical instabilities of some oxide materials at the high pressures,~\cite{Lyubutin2009, Bykova2016} which destroy the possibility of a comparison between the calculation of superexchange  interaction and experimental data.  The results of the experimental studies contain both the examples of stable   FeBO$_3$ with isostructural spin crossover \cite{Lamonova2011} at $\sim60$GPa, and chemically unstable Fe$_2$O$_3$  hematite.~\cite{Bykova2016} Further we will restrict ourselves to the stable oxide materials and assume that there are isostructural areas on the phase P/T diagram of the oxides, where the magnetic ordering is governed mainly by strong superexchange AFM
interactions in Me$-$O$-$Me with a bond angle of about $180^{\circ}$.

The aim of our work is to answer the question of how the $180^{\circ}$ superexchange interaction depends on the cation spin in transition metal oxides at the high pressure, and can simple changes in the crystal field without a spin crossover lead to a change in its nature from AFM to FM?
In terms of the realistic p-d model that include d-electrons of cation and p-electrons of oxygen the superexchange interaction arises via cation-anion p-d hopping ${t_{pd}}$  in the fourth order perturbation theory over the ${t_{pd}}$ (see for example~\cite{Eskes1993, Ohta1991, Eskes1989, Stechel1988}). Eliminating the oxygen states one can obtain the effective Hubbard model with cation-cation hopping $t\sim{{t_{pd}^2} \mathord{\left/
 {\vphantom {{t_{pd}^2} {\left( {{\varepsilon _p} - {\varepsilon _d}} \right)}}} \right.
 \kern-\nulldelimiterspace} {\left( {{\varepsilon _p} - {\varepsilon _d}} \right)}}$ and then the effective Heisenberg model may be obtained by the unitary transformation of the Hubbard model~\cite{Bulayevskii1968, Chao1977} with the superexchange interaction of the kinematic origin $J\sim{{{t^2}} \mathord{\left/
 {\vphantom {{{t^2}} U}} \right.
 \kern-\nulldelimiterspace} U}$.
 The superexchange interaction appears in a second order perturbation theory over interband hopping $t$ from the occupied low Hubbard band into the empty upper Hubbard band and back. It may be considered as result of the virtual excitation of the electron-hole pair.

We start discussing the properties of the transition metal oxides with a model of the periodic lattice of cations in d$^n$  configuration in a center of oxygen octahedra with a set of states  $\left| n \right\rangle $
with energy ${\varepsilon _n}$. The electron addition (extra electron) results in the d$^{n + 1}$  
states $\left| e \right\rangle$  with energies ${\varepsilon _e}\left( {n + 1} \right)$.  
Similar electron removal (or hole creation) involves the ${d^{n - 1}}$  states  $\left| h \right\rangle$
with energies ${\varepsilon _h}\left( {n - 1} \right)$.  
Thus, a partial contribution to the superexchange interaction involves 4 states: at site 1 creation of
the electron excite the initial state $\left| n \right\rangle$  ( we call this states  neutral) into some
$\left| e \right\rangle$ state (we call these excitation by electronic) and at site 2 the hole creation
excites the neutral state into the one of the  states  $\left| h \right\rangle$ (we call them hole).
These electon-hole excitations are virtual, after their annihilation back the final state is again
 two cations in initial d$^n$  configurations. This approach allows us to consider all partial contributions
 to the superexchange interaction including both the ground states as well as excited states in all three sectors
 of the Hilbert space: neutral ${N_0}\left( {{d^n}} \right)$, electronic ${N_ + }\left( {{d^{n + 1}}} \right)$
 and hole ${N_ - }\left( {{d^{n - 1}}} \right)$. Here we show that the sign of the partitial 
 contributions $J_{ij}^{FM}$ and $J_{ij}^{AFM}$ to  the total superexchange interaction $J_{ij} = J_{ij}^{AFM} + J_{ij}^{FM}$
 is directly independent on the cation spin $S\left( {{d^n}} \right)$, 
 but is controlled  by the spin ratio $S\left( {{d^{n - 1}}} \right) = S\left( {{d^{n + 1}}} \right)$
 (AFM interaction) or $S\left( {{d^{n - 1}}} \right) = S\left( {{d^{n + 1}}} \right) \pm 1$  (FM interaction).
 The crystal field perturbations, without a reversal of the electron spin, does not change the nature (sign) of partial contributions
 $J_{ij}^{FM}$  and $J_{ij}^{AFM}$, however  they can lead to a change in their relative magnitudes,
 as a result, to a change in the sign of superexchange parameter $J_{ij}$. 
A main factor for the comparison between the AFM and FM interactions is the type $\sigma $ or $\pi $
 overlapping orbitals involved by the partial  contributions.
These characteristics is comparable in simplicity with the well known Goodenough-Kanamori-Anderson rules,
which are used many years by scientists in the analysis of the magnetic states 
of dielectric materials.~\cite{Goodenough1963, Êànamîri1963} 
In the paper we also generalize the previous results for the superexchange interaction in iron borate under high pressure and optical pumping ~\cite{Gavrichkov2017, Gavrichkov2018} to the different transition metals oxides with magnetic ions in the d$^2$ - d$^8$ configurations.

For the readers convenience the theoretical details are placed in the Appendix below and in the main text will discuss the physical ideas.

\section{\label{sec:II} additivity properties of superexchange interaction\\}

\begin{figure}
\includegraphics{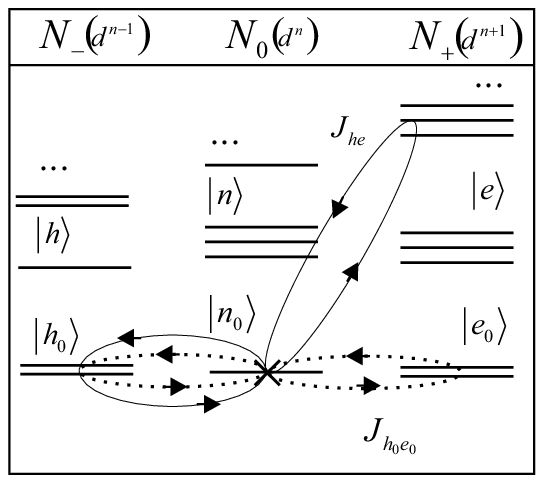}
\caption{\label{fig:1} The scheme of the superexchange interaction illustrating property of its additivity over virtual electron excitations involving all ground states  ${J_{{h_0}{e_0}}}$ (dotted line, we call this contribution the main exchange loop) and the excited electronic ${d^{n + 1}}$ contribution ${J_{{h_0}e}}$  (solid line, called the excited exchange loop).}
\end{figure}

We will work within a framework of the cell perturbation approach\cite{Gavrichkov2017} to calculate a magnitude of the superexchange, that logically fits into the LDA + GTB method to study both the electronic structure, \cite{Gavrichkov2000, Ovchinnikov2012} and the $180^\circ$ superexchange interaction in oxide materials under the pressure and optical pumping. The conclusion of our study will be some simple rules  which can help to estimate the sign of the  superexchange in the oxide materials at high pressure without complicate calculations.
At this point we will take the superexchange Hamiltonian (\ref{eq:1}) (see Appendix) as a working tool,  in structure which there is a summation over the independent contributions involving the ground $\left| {{n_0}} \right\rangle  = \left| {{{\left( {{N_0},{M_S}} \right)}_{{n_0}}}} \right\rangle$  , excited electronic $\left| {e\left( h \right)} \right\rangle  = \left| {{{\left( {{N_ \pm },{M_S}} \right)}_{e\left( h \right)}}} \right\rangle$ ($e$) and hole ($h$)  states   at  energies ${\varepsilon _{e\left( h \right)}}$  of the configuration space sectors ${N_ \pm } = n \pm 1$  for couple of the interacting magnetic cations (see. Fig.\ref{fig:1}):
\begin{eqnarray}
{\hat H_s} &=& \sum\limits_{i \ne j} {J_{ij}^{}\left( {{{\hat S}_{i{n_0}}}{{\hat S}_{j{n_0}}} - \frac{1}{4}\hat n_{i{n_0}}^{\left( e \right)}\hat n_{j{n_0}}^{\left( h \right)}} \right)},\nonumber \\
{J_{ij}} &=& \sum\limits_{he} {\frac{{{J_{ij}}\left( {h,{n_0},e} \right)}}{{\left( {2{S_h} + 1} \right)\left( {2{S_{{n_0}}} + 1} \right)}}}
\label{eq:1}
\end{eqnarray}
where $J_{ij}^{}\left( {h,{n_0},e} \right) = {{2{{\left( {t_{ij}^{{n_0}h,{n_0}e}} \right)}^2}} \mathord{\left/
 {\vphantom {{2{{\left( {t_{ij}^{{n_0}h,{n_0}e}} \right)}^2}} {{\Delta _{{n_0}he}}}}} \right.
 \kern-\nulldelimiterspace} {{\Delta _{{n_0}he}}}}$  and  ${\Delta _{{n_0}he}} = {\varepsilon _e} + {\varepsilon _h} - 2{\varepsilon _{{n_0}}}$. All definitions of the multielectron spin ${\hat S_{i{n_0}}}$   and number of quasiparticles $\hat n_{i{n_0}}^{\left( e \right)}$   operators are in the Appendix. The second contribution in Eq.(\ref{eq:1}) differs from the generally accepted method of writing the superexchange interaction and coincides with the usual form $\frac{1}{4}{\hat n_i}{\hat n_j}$  for half-filled shells, where there is electron-hole symmetry. The superexchange interaction parameter  $J_{ij}^{}$ in Eq.(1) is additive for all electronic  $\left| e \right\rangle$  and hole   $\left| h \right\rangle$ states in  sectors  ${N_ \pm }$ in Fig.\ref{fig:1} and one  is obtained in second order of cell perturbation theory over the interband contribution $\delta {\hat H_1}$ to the total Hamiltonian ${\hat H_1}$  of electron interatomic hopping:
\begin{widetext}
\begin{equation}
\delta {\hat H_1} = \sum\limits_{ij} {\hat h_{ij}^{out} = } \sum\limits_{ij} {\sum\limits_{nhe} {\left[ {t_{ij}^{el,hn}\sum\limits_\sigma  {\alpha _{i\sigma }^ + \left( {en} \right)\beta _{j\sigma }^{}\left( {hn} \right)}  + t_{ij}^{nh,le}\sum\limits_\sigma  {\beta _{i\sigma }^ + \left( {nh} \right)\alpha _{j\sigma }^{}\left( {ne} \right)} } \right]} },
\label{eq:2}
\end{equation}
\end{widetext}
that describes the creation and annihilation of the virtual electron (denoted by the operator $\alpha _{i\sigma }^ + $ ) and hole (operator  $\beta _{i\sigma }^ + $ ) pairs. Exactly the virtual excitations through the dielectric gap ${\Delta _{nhe}}$  to the conduction band and vice versa in Eq.(\ref{eq:2}) contribute to the superexchange interaction. The total multielectron Hamiltonian in the representation of the Hubbard operators~\cite{Hubbard1964} looks like $\hat H = {\hat H_0} + {\hat H_1}$, where  ${\hat H_0}$ contains all multielectron states of the involved ${d^n}$  and ${d^{n \pm 1}}$ configurations, and  ${\hat H_1}$  described all interatomic single electron hoppings (kinetic energy):

\begin{eqnarray}
&&{\hat H_0} = \sum\limits_i {\left\{ {\sum\limits_h {\left( {{\varepsilon _h} - {N_ - }\mu } \right)X_i^{hh}}  + \sum\limits_n {\left( {{\varepsilon _n} - {N_0}\mu } \right)X_i^{nn} + } } \right.} \nonumber \\
&&\left. { + \sum\limits_e {\left( {{\varepsilon _e} - {N_ + }\mu } \right)X_i^{ee}} } \right\}
\label{eq:3} \\
&&{\hat H_1} = \sum\limits_{ij} {\sum\limits_{rr'} {t_{ij}^{rr'}} X{{_i^r}^ + }} X_j^{r'}
\label{eq:4}
\end{eqnarray}
for the material with magnetic cations in arbitrary d$^n$ electron configuration. Any Hubbard operator $X_i^r = \left| p \right\rangle \left\langle q \right|$   constructed in the full and orthogonal set of eigenstates $\left| p \right\rangle$ is numerated by a pair of indexes which denotes the initial state $\left| q\right\rangle$ and the final state $\left| p \right\rangle$ of the excitation.~\cite{Hubbard1964, Ovchinnikov1997} It is more convenient to numerate each excitation by single vector index $r = ( p,q )$  (so called root vector~\cite{Zaltsev1975} that plays a role of the quasiparticle band index). Here, electronic creation operators for vector indexes  $r = (n,h)$  or $r =(e,n)$  excitations in Eq.(\ref{eq:2}) are denoted by $\beta _{i\sigma }^ + $ (${N_ - } \to {N_0}$) and  $\alpha _{i\sigma }^ + $ (${N_ 0 } \to {N_+}$) respectively.
The hopping matrix element in Eq.(\ref{eq:4}) is
\begin{equation}
t_{fg}^{rr'} = \sum\limits_{\lambda \lambda '} {t_{fg}^{\lambda \lambda '}\sum\limits_\sigma  {\left[ {\gamma _{\lambda \sigma }^*\left( r \right){\gamma _{\lambda '\sigma }}\left( {r'} \right) + \gamma _{\lambda '\sigma }^*\left( r \right){\gamma _{\lambda \sigma }}\left( {r'} \right)} \right]} }
\label{eq:5}
\end{equation}
and
\begin{equation}
{\gamma _{\lambda \sigma }}\left( r \right) = \left\langle e \right|c_{i\lambda \sigma }^ + \left| n \right\rangle  \times {\delta _{{S_{ie}},{S_{in}} \pm \left| \sigma  \right|}} \times {\delta _{{M_e},{M_n} + \sigma }}
\label{eq:6}
\end{equation}
where a root vectors  $r$ and  $r'$ run over on all possible quasiparticle excitations $(e,n)$  and   $(n,h)$ between many-electron states $\left| {n} \right\rangle$  and  $\left| {e\left( h \right)} \right\rangle$ with the energies ${\varepsilon _n}$  and  ${\varepsilon _{e\left( h \right)}}$ in the sectors ${N_0}$  and  ${N_ \pm }$ of configuration space (Fig.\ref{fig:1}). These quasiparticle excitations are described by nondiagonal elements $t_{fg}^{rr'} = t_{ij}^{nh,ne}$.  In the conventional Hubbard model there is only one such element corresponding to the excitations between lower and upper Hubbard bands. Using the results of Appendix(see Eq.(\ref{eq:A.12})), we can represent the exchange parameter for a pair of interacting spins ${S_{i{n_0}}} = {S_{j{n_0}}}$ in the form of Eq.(\ref{eq:7}):
\begin{equation}
{J_{ij}} = J_{ij}^{AFM} + J_{ij}^{FM}.
\label{eq:7}
\end{equation}
This equality and its relationship with the spin ${S_{h\left( e \right)}}$  at the states  $\left| {h\left( e \right)} \right\rangle$  was obtained in the works~\cite{Gavrichkov2017, Gavrichkov2018} for iron borate and also was firstly briefly mentioned in the works.~\cite{lrkhin1993, Irkhin1994} The virtual electron interband $\left( {{n_0},e} \right)$  and  $\left( {{n_0},h} \right)$  hoppings correspond to only one of contributions in the sum ${J_{ij}} = \sum\limits_{he} {{J_{ij}}\left( {h,{n_0},e} \right)}$, and any contribution ${J_{he}} = \sum\limits_{ij} {J_{ij}^{}\left( {h,{n_0},e} \right)}$   can be represented by a double loop or the so-called exchange loop, marked by the same line (solid or dashed).  In Fig.\ref{fig:1} the contributions  ${J_{he}}$ is illustrated by double exchange loops with the arrows which connect the ground state  of the magnetic ions $\left| {{n_0}} \right\rangle$ with all ground $\left| {h_0\left( e_0 \right)} \right\rangle$ and excited $\left| {h\left( e \right)} \right\rangle$  states.

\section{\label{sec:III} Rules for a sign of different contributions to the superexchange \\}

The new result of this paper is the classification of different contrubutions by the relation between spins ${S_h}$  and ${S_e}$.  If in exchange loop ${S_h} = {S_e} \pm 1$  it will be FM contribution, in the other case  ${S_h} = {S_e}$ it is AFM contribution.  These two relations exhaust all possible interrelations between spins for all nonzero contributions, i.e. in any other case, the contribution to superexchange from this pair of states $\left| h \right\rangle $ and  $\left| e \right\rangle $ is simply not available. The sign of the total exchange interaction  (FM or AFM) depends on a relationship between relative magnitudes of the contributions.  The main difficulty is a great number of excited states in ${N_ \pm }$  sectors of the configuration space. Due to the smallest denominator ${\Delta _{{n_0}{h_0}{e_0}}}$  in the superexchange (\ref{eq:1}),  the main exchange loop involving ground  $\left| {{h_0}\left( {{e_0}} \right)} \right\rangle$ states can form a dominant ${J_{{h_0}{e_0}}}$  contribution. However, the contributions  ${J_{he}}$ from the excited states  $\left| {h\left( e \right)} \right\rangle$ in   ${N_ \pm }$ sectors can compete with  the main exchange loop  due to the dominant nominator, if the excited exchange loop occurs by overlapping of states with ${e_g}$ symmetry, and the main exchange loop is formed by $\pi$ bonding despite not the smallest denominator ${\Delta _{{n_0}he}}$  in Eq.(\ref{eq:1}).  The problem is that without complicated numeric calculation taking into account all hopping integrals (\ref{eq:4}), it is difficult to obtain the final answer about the magnitude and sign of the superexchange interaction. For example, such numerical calculations have been carried out for   La$_2$CuO$_4$ with a configuration $d^9$, where  a number of the contributions exceeds ten ones.~\cite{Gavrichkov2016, Gavrichkov2008} We will give a qualitative criterion that takes into account both factors  in the case both $\sigma$  or   $\pi$ overlapping in the Hamiltonian (\ref{eq:2})  (where $t_{ij}^{el,hn}$  hopping is obtained by the mapping of the multiband p-d model, which includes integral for $\sigma$  or   $\pi$ overlapping), and the energy gap  ${\Delta _{nhe}}$ in the arbitrary exchange loop ${J_{he}}$. The minimal gap ${\Delta _{{n_0}{h_0}{e_0}}}$  just coincide with a dielectric gap $E_g$ in the oxide materials.  After comparing calculated sign of the superexchange constant for magnetic ions in the electron configurations d$^2$ - d$^8$  with experimental data, we found that in most cases there is no need to sum over all possible virtual hoppings (or exchange loops), it is enough to establish the criterion in form:\\
1. For the  $\sigma$ overlapping $e_g$ states corresponding to contribution ${J_{{h_0}{e_0}}}$, the sign of superexchange is controlled by the virtual electron excitations with participate of the ground  $\left| {{h_0}\left( {{e_0}} \right)} \right\rangle$ states and minimal magnitude of the energy gap ${\Delta _{{n_0}{h_0}{e_0}}}\sim{E_g}$ .  These excitations involved to the main exchange loop is pictured in Fig.\ref{fig:1}  by a dashed line.\\
2.  In the case of $\pi$ overlapping ${t_{2g}}$ states for the virtual electron excitations involving only the ground  $\left| {{h_0}\left( {{e_0}} \right)} \right\rangle$ states the sign of superexchange is controlled by not  the main exchange loop, but the virtual electron excitations (exchange loop) involving the excited states with the $\sigma$  overlapping $e_g$ states. These virtual excitations are pictured in Fig.\ref{fig:1} by solid line. If such exchange loops is absent, the sign of superexchange is controlled still by the main loop with the  $\pi$ overlapping.\\
Here, the $\sigma$  overlapping have the priority.  Indeed, the superexchange interaction is proportional to the fourth degree of the overlapping integral ${I_{\sigma \left( \pi  \right)}} = \rho \left( {\left| {\Delta R} \right|} \right){\chi _{\sigma \left( \pi  \right)}}$   between the electron states of the anion and the magnetic cation, where the radial part $\rho \left( {\left| {\Delta R} \right|} \right)$  depends only on the anion-cation distance $\Delta R$, and the angular part ${\chi _{\sigma \left( \pi  \right)}}$ depends on the angular distribution of  the anions. The squared ratio ${\left( {{{I_\pi ^{}} \mathord{\left/
 {\vphantom {{I_\pi ^{}} {I_\sigma ^{}}}} \right.
 \kern-\nulldelimiterspace} {I_\sigma ^{}}}} \right)^2}$ of the overlapping integrals for $e_g$  and $t_{2g}$ states involved in the superexchange through $\sigma$  and $\pi$  coupling in the same octahedral complexes is the following relation: ${\left( {{{I_\pi ^{}} \mathord{\left/
 {\vphantom {{I_\pi ^{}} {I_\sigma ^{}}}} \right.
 \kern-\nulldelimiterspace} {I_\sigma ^{}}}} \right)^2} = {\left( {{{\chi _\tau ^{}} \mathord{\left/
 {\vphantom {{\chi _\tau ^{}} {\chi _\sigma ^{}}}} \right.
 \kern-\nulldelimiterspace} {\chi _\sigma ^{}}}} \right)^2} = {1 \mathord{\left/
 {\vphantom {1 3}} \right.
 \kern-\nulldelimiterspace} 3}$.  Thus the fourth degree gives ratio of matrix elements $\sim 0.1$,  i.e., competition between the contributions with a participation of virtual ${t_{2g}}$  electron hopping and the one through $\sigma$ coupling  is possible, when the denominator energy ${\Delta _{nhe}}$  for excited loop $ J_{he}$ is no more than 9 times higher in energy than the main loop energy  ${\Delta _{{n_0}{h_0}{e_0}}}$. Otherwise the $\sigma$ type contribution from exchange loop  is  dominant. In case of several competing contributions simple calculations of the multielectron energies below and above the spin crossover at the high pressure~\cite{Ovchinnikov2008} can be used to compare energy denominators of the AFM and FM contributions given by Eq.(\ref{eq:7}). Some examples will be given in the next section for oxide materials with d$^7$  and d$^5$  cations.\\

\section{\label{sec:IV} Superexchange in oxides with cations in $d^7$ and  $d^5$ configurations \\}

\begin{figure}
\includegraphics{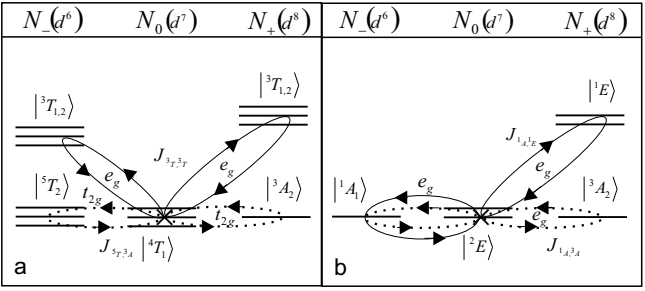}
\caption{\label{fig:2} Scheme of the $180^\circ$ superexchange interaction in CoO: (a) at the ambient pressure,  where AFM interaction  is controlled by the contribution from the  exchange loop ${J_{{}^{{}^3T,{}^3T}}}$ with the excited states $\left| {{}^3{T_{1,2}}} \right\rangle$   and $\sigma$  overlapping. The contribution ${J_{{}^{{}^5T,{}^3A}}}$  from the main exchange loop ${J_{{}^{{}^1A,{}^3A}}}$ with $\pi$ overlapping is showed by a dotted line; (b) under high pressure,  where FM character is controlled by the   main exchange loop ${J_{{}^{{}^1A,{}^3A}}}$   with the $\sigma$ overlapping. The AFM contribution from the  exchange loop ${J_{{}^{{}^1A,{}^1E}}}$ with participation of excited states $\left| {{}^1E} \right\rangle$  has a large denominator. }
\end{figure}

Let us show, using the example of oxide materials  CoO and Ni$_2$O$_3$  with Ni$^3+$, Co$^2+$ cations in the  d$^7$ electron configuration under high pressure, how our rules work.  The energy of the neutral  $\left| n \right\rangle$ (d$^7$) states and electronic $\left| e\right\rangle$  (d$^8$) and hole  $\left| h \right\rangle$ (d$^6$)  states at  the ambient pressure are shown in Fig.\ref{fig:2}(a).  From the main exchange loop with $\pi$  overlapping our rules results in the FM sign of the contribution ${J_{{}^{{}^5T,{}^3A}}}$. Competing AFM contribution is the  exchange loop ${J_{{}^{{}^3T,{}^3T}}}$ with the excited states  $\left| {{}^3{T_{1,2}}} \right\rangle$ and $\sigma$  overlapping. Below we will check our rules by direct calculation for the main exchange loop.  To derive the FM contribution ${J_{{}^{{}^5T,{}^3A}}}$  using angular momentum addition rules, we introduce the creation operators $\beta _{i\sigma }^ + \left( {{n_0},{h_0}} \right)$   for  ${N_ - } \leftrightarrow {N_0}$ hole quasiparticles by Eq.(\ref{eq:7}) and   $\alpha _{i\sigma }^ + \left( {{e_0},{n_0}} \right)$ for ${N_0} \leftrightarrow {N_ + }$  electron quasiparticles by Eq.(\ref{eq:8}).~\cite{Gavrichkov2017}
\begin{widetext}
\begin{eqnarray}
&&- \beta _{i \uparrow }^ +  = \sqrt {\frac{1}{5}} X_i^{\frac{3}{2},1} + \sqrt {\frac{2}{5}} X_i^{\frac{1}{2},0} + \sqrt {\frac{3}{5}} X_i^{ - \frac{1}{2}, - 1} + \sqrt {\frac{4}{5}} X_i^{ - \frac{3}{2}, - 2},
\beta _{i \downarrow }^ +  = \sqrt {\frac{4}{5}} X_i^{\frac{3}{2},2} + \sqrt {\frac{3}{5}} X_i^{\frac{1}{2},1} + \sqrt {\frac{2}{5}} X_i^{ - \frac{1}{2},0} + \sqrt {\frac{1}{5}} X_i^{ - \frac{3}{2}, - 1} \nonumber \\
&&- \alpha _{i \uparrow }^ + \left( {{}^3{A_2},{}^4T} \right) = \sqrt {\frac{1}{4}} X_i^{1,\frac{1}{2}} + \sqrt {\frac{1}{2}} X_i^{0, - \frac{1}{2}} + \sqrt {\frac{3}{4}} X_i^{ - 1, - \frac{3}{2}},
\alpha _{i \downarrow }^ + \left( {{}^3{A_2},{}^4T} \right) = \sqrt {\frac{3}{4}} X_i^{1,\frac{3}{2}} + \sqrt {\frac{1}{2}} X_i^{0,\frac{1}{2}} + \sqrt {\frac{1}{4}} X_i^{ - 1, - \frac{1}{2}}
\nonumber \\
\label{eq:8}
\end{eqnarray}
\end{widetext}
Working further in framework of the cell perturbation theory,  we obtain in the second order the FM contribution  ${J_{{}^{{}^5T,{}^3A}}}$ from the main exchange loop in Fig.\ref{fig:2}  with the $\pi$ overlapping:
\begin{equation}
{J_{{}^{{}^5T,{}^3A}}} =  - \sum\limits_{i \ne j} {\frac{{J_{ij}^{}\left( {{}^5T,{}^3A} \right)}}{{\left( 5 \right)\left( {{3 \mathord{\left/
 {\vphantom {3 2}} \right.
 \kern-\nulldelimiterspace} 2}} \right)}}\left( {{{\hat S}_{i{n_0}}}{{\hat S}_{j{n_0}}} + \frac{1}{4}\hat n_{i{n_0}}^{\left( e \right)}\hat n_{j{n_0}}^{\left( h \right)}} \right)}
 \label{eq:9}
\end{equation}
where ${S_{i{n_0}}} = \frac{3}{2}$, $\hat S_{i{n_0}}^ +  =  - 5\beta _{i \uparrow }^ + \beta _{i \downarrow }^{} =  - 4{\alpha _{i \downarrow }}\alpha _{i \uparrow }^ +$, $\hat S_{i{n_0}}^z =  - 5\sum\limits_\sigma  {\eta \left( \sigma  \right)\beta _{i\sigma }^ + {\beta _{i\sigma }}}  =  - 4\sum\limits_\sigma  {\eta \left( \sigma  \right){\alpha _{i\sigma }}\alpha _{i\sigma }^ + }$, and also $\hat n_{i{n_0}}^{\left( e \right)} = 5\sum\limits_\sigma  {\beta _{i\sigma }^ + {\beta _{i\sigma }}}$ and $\hat n_{j{n_0}}^{\left( h \right)} = 4\sum\limits_\sigma  {{\alpha _{j\sigma }}\alpha _{j\sigma }^ + }$  are the number of electron and hole quasiparticles involved in the superexchange interaction.  According to a second point of the criterion the FM contribution competes with the AFM  ${J_{{}^{{}^3T,{}^3T}}}$ contribution:
\begin{equation}
{J_{{}^{{}^3T,{}^3T}}} = \sum\limits_{i \ne j} {\frac{{J_{ij}^{}\left( {{}^3T,{}^3T} \right)}}{{\left( 3 \right)\left( {{3 \mathord{\left/
 {\vphantom {3 2}} \right.
 \kern-\nulldelimiterspace} 2}} \right)}}\left( {{{\hat S}_{i{n_0}}}{{\hat S}_{j{n_0}}} - \frac{1}{4}\hat n_{i{n_0}}^{\left( e \right)}\hat n_{j{n_0}}^{\left( h \right)}} \right)}
\label{eq:10}
\end{equation}
from the virtual hoppings of ${e_g}$  electrons with participation of  the  states $\left| {{}^3{T_{1,2}}} \right\rangle$  and $\sigma$ overlapping (see Fig.\ref{fig:2}(a)). Similarly to Eqs.(\ref{eq:7}) and (\ref{eq:8}), new  ${\alpha'}_{i \sigma }^+$  and ${\beta'}_{i \sigma }^+$  quasiparticles involved in this superexchange are given by the expression:
\begin{widetext}
\begin{eqnarray}
&&{\beta'}_{i \uparrow }^ + \left( {{}^4T,{}^3T} \right) = X_i^{\frac{3}{2},1} + \sqrt {\frac{2}{3}} X_i^{\frac{1}{2},0} + \sqrt {\frac{1}{3}} X_i^{ - \frac{1}{2}, - 1},
{\beta'}_{i \downarrow }^ + \left( {{}^4T,{}^3T} \right) = \sqrt {\frac{1}{3}} X_i^{\frac{1}{2},1} + \sqrt {\frac{2}{3}} X_i^{ - \frac{1}{2},0} + X_i^{ - \frac{3}{2}, - 1}; \nonumber \\
&&- {\alpha'}_{i \uparrow }^ + \left( {{}^3T,{}^4T} \right) = \sqrt {\frac{1}{4}} X_i^{1,\frac{1}{2}} + \sqrt {\frac{2}{4}} X_i^{0, - \frac{1}{2}} + \sqrt {\frac{3}{4}} X_i^{ - 1, - \frac{3}{2}},
{\alpha'}_{i \downarrow }^ + \left( {{}^3T,{}^4T} \right) = \sqrt {\frac{3}{4}} X_i^{1,\frac{3}{2}} +\sqrt {\frac{1}{2}} X_i^{0,\frac{1}{2}} + X_i^{ - 1, - \frac{1}{2}} \nonumber \\
\label{eq:11}
\end{eqnarray}
\end{widetext}
Here: $\hat S_{i{n_0}}^ +  = 3{\beta'}_{i \uparrow }^ + {\beta'}_{i \downarrow }^{} =  - 4{{\alpha'}_{i \downarrow }}{\alpha'}_{i \uparrow }^ +$, $\hat S_{i{n_0}}^z = 3\sum\limits_\sigma  {\eta \left( \sigma  \right){\beta'}_{i\sigma }^ + {{\beta'}_{i\sigma }}}  =  - 4\sum\limits_\sigma  {\eta \left( \sigma  \right){{\alpha '}_{i\sigma }}{\alpha '}_{i\sigma }^ + }$  and  $\hat n_{i{n_0}}^{\left( e \right)} = 3\sum\limits_\sigma  {{\beta '}_{i\sigma }^ + {{\beta '}_{i\sigma }}}$, $\hat n_{i{n_0}}^{\left( h \right)} = 4\sum\limits_\sigma  {{{\alpha '}_{i\sigma }}{\alpha '}_{i\sigma }^ + }$
Calculation of energies of the different states below and above spin crossover allows us to obtain the energy denominators for the different contributions to superexchange interaction. For the main exchange loop ${J_{{}^{{}^5T,{}^3A}}}$ in Fig.\ref{fig:2}(a) the value ${\Delta _{{n_0}he}} = U - {J_H}$, where $U$  is the intra-atomic Coulomb matrix element (Hubbard parameter) and ${J_H}$  is the Hund exchange coupling, both  $U$ and  ${J_H}$  are positive. For the contribution  from exchange loop ${J_{{}^{{}^3T,{}^3T}}}$, ${\Delta _{{n_0}he}} = {\varepsilon _e} + {\varepsilon _h} - 2{\varepsilon _{{n_0}}} = U + {J_H}$.  At the typical magnitudes $U = 6eV$  and   ${J_H} = 1eV$ the ration of denominators is $5/8$, and the ratio of numerators is $9/1$. It proves the dominant AFM contribution below spin crossover. With increasing  pressure there is the spin crossover in   configuration d$^7$. The pressure enter in the crystal field parameter $10Dq$ that linearly increases with the pressure: below spin crossover at the ambient pressure when  $10Dq < 2{J_H}$ the cation $C{o^{3 + }}$ is at the HS state, and $\left| {{n_0}} \right\rangle  = \left| {{}^4{T_1}} \right\rangle$, $\left| {{h_0}} \right\rangle  = \left| {{}^5{T_2}} \right\rangle $, $\left| {{e_0}} \right\rangle  = \left| {{}^3{A_1}} \right\rangle$  (see Fig.\ref{fig:2}(a)).  Above spin crossover at $10Dq > 2{J_H}$  the cation $C{o^{3 + }}$  is at the LS  state $\left| {{n_0}} \right\rangle  = \left| {{}^2E} \right\rangle$, and $\left| {{h_0}} \right\rangle  = \left| {{}^1A} \right\rangle$ (see Fig.\ref{fig:2}(b)).~\cite{Ovchinnikov2008} Thus,  the ground $\left| {{n_0}} \right\rangle$ and hole $\left| {{h_0}} \right\rangle$ states    the superexchange interactions in the cobalt monoxide  under high pressure is changed.  The main exchange loop ${J_{{}^{{}^1A,{}^3A}}}$  with the  $\sigma$ overlapping should be FM according our rules.
\begin{equation}
{J_{{}^{{}^1A,{}^3A}}} =  - \sum\limits_{i \ne j} {\frac{{J_{ij}^{}\left( {{}^1A,{}^3A} \right)}}{2}\left( {{{\hat S}_{i{n_0}}}{{\hat S}_{j{n_0}}} + \frac{1}{4}\hat n_{i{n_0}}^{\left( e \right)}\hat n_{j{n_0}}^{\left( h \right)}} \right)}
\label{eq:12}
\end{equation}
The AFM contribution   from the exchange loop with the excited states   has the large denominator    than the FM one (Fig.\ref{fig:2}b).
\begin{equation}
{J_{{}^{{}^1A,{}^1E}}} = \sum\limits_{i \ne j} {\frac{{J_{ij}^{}\left( {{}^1A,{}^1E} \right)}}{2}\left( {{{\hat S}_{i{n_0}}}{{\hat S}_{j{n_0}}} - \frac{1}{4}\hat n_{i{n_0}}^{\left( e \right)}\hat n_{j{n_0}}^{\left( h \right)}} \right)}
\label{eq:13}
\end{equation}
These conclusions can be obtained analogously to the previous Eq.(\ref{eq:9}) and Eq.(\ref{eq:10}), starting from building operators $\beta _{i\sigma }^ +$, $\alpha _{i\sigma }^ +$   and ${\beta '}_{i\sigma }^ +$,   ${\alpha '}_{i\sigma }^ +$ of the  quasiparticles and finishing with derivation of the Eqs.(\ref{eq:12}) and (\ref{eq:13}). We have to compare the energy denominators.  For FM   contribution ${J_{{}^{{}^1A,{}^3A}}}$,  the energy ${\Delta _{{}^1A{}^3A}} = \varepsilon \left( {{}^1A,{d^6}} \right) + \varepsilon \left( {{}^3A,{d^8}} \right) - \varepsilon \left( {{}^2E,{d^7}} \right) = U - {J_H}$ and ${\Delta _{{}^1A{}^1E}} = U$. Taking into account that all contributions have the same  $\sigma$  bonding, we came to conclusion that resulting interaction in the LS state for materials with the cations in d$^7$ configuration will be FM.

Let's compare our conclusions with the results for iron borate FeBO$_3$  at the high pressure. Under pressure $P\sim60GPa$  in the iron borate  with  cations Fe$^{3+}$ in the    configuration d$^5$ the spin crossover  $\left| {{}^6{A_1}} \right\rangle  \to \left| {{}^2{T_2}} \right\rangle$ occurs at $10Dq = 3{J_H}$. Given above criterion tells us that the sign of the exchange interaction in iron borate is changed from AFM to FM with increasing pressure in agreement with direct calculations.~\cite{Gavrichkov2018}. This conclusion is also valid for another oxide materials with  cations in the configuration d$^5$ and octahedral environment.
\begin{figure}
\includegraphics{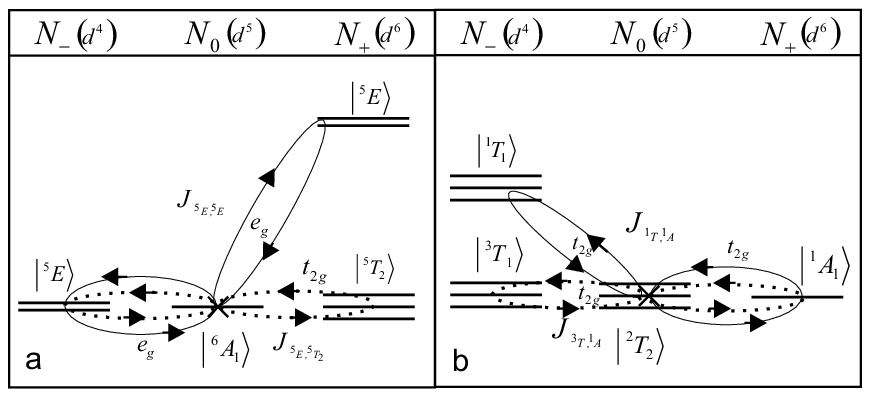}
\caption{\label{fig:3} Scheme of the $180^\circ$ superexchange interaction in FeBO$_3$: (a)  at the ambient pressure,  where the  main exchange loop ${J_{{}^{{}^5E,{}^5{T_2}}}}$ has a zero contribution because of zero overlapping, and the  $\sigma$  overlapping exchange loops ${J_{{}^{{}^5E,{}^5E}}}$   result in AFM  contribution only; (b) under high pressure, where both contributions ${J_{{}^{{}^3T,{}^1A}}}$ (FM) and  ${J_{{}^{{}^1T,{}^1A}}}$ (AFM) are proportional to   $\pi$ overlaping. The FM contribution  ${J_{{}^{{}^3T,{}^1A}}}$  dominates. }
\end{figure}

\begin{table*}
\caption {The examples of transition metals oxides with calculated sign of $180^{\circ}$ superexchange interactions (in 3 and  5 columns), and also the magnetic ordering below and above the spin crossover (in 4 and 6 columns). The notations (ex) and (gr)  indicates the nature of the main contribution to the superexchange: (ex) is the exchange loop involving excited states, (gr) is the main exchange loop.}
\label{tab:1}
\begin{ruledtabular}
\begin{tabular}{cccccc}
 \footnotesize Cation and electron             &\footnotesize Oxides            &\footnotesize Superexchange &\footnotesize Ambient pressure    &\footnotesize Superexchange  &\footnotesize High pressure\\
\footnotesize  configuration                   &\footnotesize                               &\footnotesize below spin crossover      &\footnotesize (experiment)  &\footnotesize above spin crossover &\footnotesize (experiment)\\
\hline \\
\footnotesize d$^2$, Cr$^{4+}$, &\footnotesize CrO$_2$      &\footnotesize  $J_{{}^2T,{}^4A}^{FM}$(gr) &\footnotesize FM, ${T_C} = 90K$ &\footnotesize no crossover,  &\footnotesize FM up to \\
\footnotesize $S_{n_0}=1$  &\footnotesize  &\footnotesize &\footnotesize &\footnotesize $J_{{}^2T,{}^4A}^{FM}$ (gr),  &\footnotesize P=56GPa,~\cite{Kuznetsov2006}\\
\footnotesize &\footnotesize  &\footnotesize &\footnotesize &\footnotesize $S_{n_0}=1$  &\footnotesize  \\
\hline \\
                                   \footnotesize d$^3$, Cr$^{3 + }$,  &\footnotesize        LaCrO$_3$               &\footnotesize $J_{{}^3T,{}^3T}^{AFM}$(ex)     &\footnotesize AFM, ${T_N} = 298K$ \cite{Zhou2011}  &\footnotesize no crossover &\footnotesize AFM, $T_N$ increases \\
                                   \footnotesize ${S_{{n_0}}} = \frac{3}{2}$ &                         &\footnotesize     &\footnotesize &\footnotesize $J_{{}^3T,{}^3T}^{AFM}$(gr), &\footnotesize with a pressure up to \\
                                   \footnotesize  &	        			          &\footnotesize    &\footnotesize &\footnotesize  ${S_{{n_0}}} = \frac{3}{2}$& \footnotesize 380K at P=6.5 GPa~\cite{Zhou2011}  \\
                                   \footnotesize &	                        & \footnotesize    &\footnotesize    &\footnotesize &\footnotesize \\
\hline \\
\footnotesize d$^4$, Fe$^{4 + }$, Mn$^{3 + }$,	 &\footnotesize LaMnO$_3$          &\footnotesize $J_{{}^{{}^4A,{}^6A}}^{FM}(gr)$                 &\footnotesize AFM, with FM planes &\footnotesize crossover is expected  &\footnotesize AFM, ${T_N} = 152K$ \\
\footnotesize $S_{n_0}=2$	 &\footnotesize           &\footnotesize                  &\footnotesize ${T_N} = 140K$,
~\cite{Zhou2002}&\footnotesize to the LS state, &\footnotesize at the pressure \\
\footnotesize	 &\footnotesize    &\footnotesize                 &\footnotesize &\footnotesize $J_{{}^4A,{}^2T}^{FM}$(gr),
&\footnotesize P=2 GPa. FM above \\
\footnotesize	 &\footnotesize      &\footnotesize	            &\footnotesize &\footnotesize $S_{n_0}=1$ &\footnotesize the spin crossover \\
\footnotesize &\footnotesize	       &\footnotesize                  &\footnotesize &\footnotesize  &\footnotesize is predicted. \\

\hline \\
\footnotesize d$^5$, Fe$^{3 + }$, Mn$^{2 + }$	 &\footnotesize FeBO$_3$,  	       &\footnotesize $J_{{}^{{}^5E,{}^5E}}^{AFM}$(ex)                 &\footnotesize AFM, ${T_N} = 348K$\cite{Gavriliuk2004} &\footnotesize spin crossover, &\footnotesize ${T_{N\left( C \right)}} = 50K^*$, \\
\footnotesize ${S_{{n_0}}} = \frac{5}{2}$  &\footnotesize  (Fe$_2$O$_3$, MnO        &\footnotesize                &\footnotesize &\footnotesize $J_{{}^{{}^3T,{}^1A}}^{FM}$(gr)
 &\footnotesize at P=49 GPa, \\
\footnotesize	 &\footnotesize    &\footnotesize                 &\footnotesize &\footnotesize ${S_{{n_0}}} = \frac{1}{2}$ &\footnotesize FM above the \\
\footnotesize  &\footnotesize    &\footnotesize
            &\footnotesize &\footnotesize  &\footnotesize  spin crossover is\\
\footnotesize &\footnotesize    &\footnotesize                &\footnotesize &\footnotesize
&\footnotesize predicted. \\

\hline \\
\footnotesize d$^6$, Fe$^{2 + }$, Co$^{3 + }$	 &\footnotesize Mg$_{1 - x}$Fe$_x$O,	       &\footnotesize $J_{{}^4T,{}^4T}^{AFM}$(ex)          &\footnotesize AFM, ${T_N} = 37K$~\cite{Lyubutin2012} &\footnotesize spin crossover to &\footnotesize non magnetic \\
\footnotesize ${S_{{n_0}}} = 2$  &\footnotesize   (LaCoO$_3$)       &\footnotesize                &\footnotesize &\footnotesize nonmagnetic  state
 &\footnotesize above P=55 GPa~\cite{Lyubutin2012, Lyubutin2013} \\
\footnotesize	 &\footnotesize    &\footnotesize                 &\footnotesize &\footnotesize with ${S_{{n_0}}} = 0$ &\footnotesize \\

\hline \\
\footnotesize d$^7$, Co$^{2 + }$, Ni$^{3 + }$,	 &\footnotesize CoO,    &\footnotesize $J_{{}^{{}^3T,{}^3T}}^{AFM}$(ex)   &\footnotesize AFM, ${T_N} = 290K$\cite{Jauch2001}
&\footnotesize spin crossover &\footnotesize spin crossover\\
\footnotesize ${S_{{n_0}}} = \frac{3}{2}$	 &\footnotesize (La$_2$CoO$_4$, LaNiO$_3$)   &\footnotesize   &\footnotesize  &\footnotesize is expected,
 &\footnotesize observed at \\
\footnotesize &\footnotesize    &\footnotesize   &\footnotesize &\footnotesize $J_{{}^{{}^1A,{}^3A}}^{FM}$(gr),
&\footnotesize P=80-90 GPa~\cite{Atou2004,Guo2002} \\
\footnotesize &\footnotesize    &\footnotesize   &\footnotesize &\footnotesize  ${S_{{n_0}}} = \frac{1}{2}$ &\footnotesize \\
\hline \\

\footnotesize d$^8$, Ni$^{2+}$, Cu$^{3+}$ &\footnotesize NiO   &\footnotesize $J_{{}^2E,{}^2E}^{AFM}$(ex)   &\footnotesize AFM, ${T_N} = 525K$
&\footnotesize no spin crossover, &\footnotesize no spin crossover\\
\footnotesize ${S_{{n_0}}} = 1$	 &\footnotesize    &\footnotesize   &\footnotesize &\footnotesize $J_{{}^2E,{}^2E}^{AFM}$,
 &\footnotesize observed up to\\
\footnotesize &\footnotesize    &\footnotesize   &\footnotesize &\footnotesize ${S_{{n_0}}} = 1$
&\footnotesize P=220 GPa~\cite{Lyubutin2009,Rinaldi-Montes2016} \\

\end{tabular}
\end{ruledtabular}
*The critical temperature ${T_{N\left( C \right)}}$  of magnetic ordering in iron borate FeBO$_3$  at the higher pressure was measured by Mossbauer spectroscopy,~\cite{Lyubutin2009, Gavriliuk2004} however,  this method cannot distinguish the nature (FM or AFM) of the magnetic ordering. Up to now there is no experimental data on the magnetic ordering in the LS state of  FeBO$_3$ or any other materials with d$^5$  cations.
\end{table*}

At the ambient pressure FM contributions from the exchange loops are missing (Fig.\ref{fig:3}(a)). The AFM  superexchange interaction is caused by the    contribution ${J_{{}^{{}^5E,{}^5E}}}$ from the $\sigma$ bonding exchange loop with the excited   $\left| e \right\rangle$ states. The calculation of the energy denominator is ${\Delta _{{}^5E{}^5T}} = U - 10Dq + 4{J_H}$.  Thus, the AFM exchange interaction at the ambient pressure may be estimated as ${J_{{}^5E{}^5T}} \approx {{t_\sigma ^2} \mathord{\left/
 {\vphantom {{t_\sigma ^2} {\left( {U + {J_H}} \right)}}} \right.
 \kern-\nulldelimiterspace} {\left( {U + {J_H}} \right)}}$.  Crystal field increases with pressure, and at the critical pressure $10Dq\left( {{P_c}} \right) = 3J{}_H$  there is spin crossover $\left| {{}^6{A_1}} \right\rangle  \to \left| {{}^2{T_2}} \right\rangle$.  Above the spin crossover, the nature of the FM superexchange interaction is obtained from the competition of FM(${J_{{}^{{}^3T,{}^1A}}}$) and AFM (${J_{{}^{{}^1T,{}^1A}}}$)  loops with the same type of  $\pi$ overlapping, where the FM contribution prevails (see Fig.\ref{fig:3}(b) due to the smaller magnitude of the energy gap ${\Delta _{{n_0}he}}$. We can estimate the competing FM and AFM by calculation of their energy denominators. For the main FM exchange loop (dotted lines in Fig.\ref{fig:3}(b) the energy ${\Delta _{{}^5T{}^1A}} \approx U - {J_H}$,  and for  the excited AFM loop (solid lines in Fig.\ref{fig:3}(b) the energy ${\Delta _{{}^1T{}^1A}} \approx U$. That is why the FM contribution dominates. Nevertheless the AFM one strongly reduced the total FM superexchange interaction, that can be estimated as
\begin{equation}
{J_{FM}} = {J_{{}^5E{}^5T}} + {J_{{}^5E{}^5T}} \approx \frac{{t_\pi ^2}}{{U - {J_H}}} - \frac{{t_\pi ^2}}{U} = \frac{{t_\pi ^2}}{{U - {J_H}}}\frac{{{J_H}}}{U}
\label{eq:14}
\end{equation}
Thus, spin crossover in oxide materials with ${d^5}$ cations not only changes the sign of exchange interaction, but also reduces its amplitude by the factor ${{{J_H}} \mathord{\left/
 {\vphantom {{{J_H}} {U <  < 1}}} \right.
 \kern-\nulldelimiterspace} {U <  < 1}}$  .

\section{\label{sec:V} Superexchange in oxides with cations in other electron configurations  \\}

 Now, we can obtain the nature(FM or AFM) of the superexchange interaction for oxide materials with ${d^2}$ - ${d^8}$ cations under pressure, below and above the spin crossover in Tab.1, and also compare one with experimental data, where it is possible. In the oxide materials with another ${d^n}$  ions, where $n = 2,3$, spin crossover is not possible, and ground states  $\left| {{}^3{T_1}} \right\rangle$ and $\left| {{}^4{A_2}} \right\rangle $ is stable under high pressure.

 d$^2$. Chromium dioxide CrO$_2$, where chrome cation Cr$^{4 + }$ has   configuration d$^2$ with spin ${S_{{n_0}}} = 1$, is the example of FM contribution $J_{{}^2T,{}^4A}^{}$  from the main  exchange loop involving the ground states of ${t_{2g}}$  cation with the $\pi$  overlapping at an arbitrarily pressure. FM ordering in chromium dioxide is known experimentally and persists in orthorhombic phase of the chromium dioxide  up to P=56Gpa.~\cite{Kuznetsov2006}

 d$^3$. For chromium oxide LaCrO$_3$  with  cations Cr$^{3 + }$ at the ground state $\left| {{}^4{A_2}} \right\rangle$  is stable under pressure, and the dominant AFM contribution is given by the exchange loop with the ground     state $\left| {{h_0}} \right\rangle  = \left| {{}^3T} \right\rangle$ in the hole   configuration d$^2$ and the excited   state $\left| n \right\rangle  = \left| {{}^3T} \right\rangle$ in the electron  ${d^4}$ configuration. Under high pressure when    $10Dq(P) > 3{J_H}$ the crossover stabilizes the triplet $\left| {{n_0}} \right\rangle  = \left| {{}^3T} \right\rangle$. The AFM sign of the exchange interaction does not change, but the same    interaction $J_{{}^3T,{}^3T}^{}$ is described by the main exchange loop and its value becomes larger.

 d$^4$. In manganite LaMnO$_3$  at the ambient pressure with  cations Mn$^{3 + }$ at the ground HS state $\left| {{n_0}} \right\rangle  = \left| {{}^5E} \right\rangle$,  the $\sigma$ overlapping main loop  $J_{{}^4A,{}^6A}^{}$ results in the FM interaction.  Under high pressure $10Dq(P) > 3{J_H}$  when the   cations $M{n^{3 + }}$ are in the intermediate spin state $\left| {{n_0}} \right\rangle  = \left| {{}^3T} \right\rangle$, and all superexchange interactions results from the  $\pi$ bonding. The main exchange loop provides the FM interaction $J_{{}^4A,{}^2T}^{}$, with the energy denominator ${\Delta _{{}^4A{}^2T}} = U - {J_H}$, while exchange via  excited states gives the AFM contribution $J_{{}^2T,{}^2T}^{}$ with ${\Delta _{{}^2T{}^2T}} = U + {J_H}$, and the total superexchange interaction has the FM sign. It should be emphasized that in this study we consider the crystals with cations in the octahedral oxygen environment. When we compare our conclusions about the FM interaction with the magnetic state of manganite, we find the disagreement with its AFM ordering at the ambient pressure. Nevertheless this AFM ordering consists of the FM ab planes that are AFM coupled. This disagreement is probably related to the dependence of the magnetic ordering on the type of orbital ordering in the oxide material with Jahn - Teller  cations Mn$^{3 + }$.~\cite{Kugel1982, Streltsov2017} With increasing pressure, the spin crossover in   is accompanied by the transition of the   cation Mn$^{3 + }$ from the HS   Jahn - Teller state $\left| {{}^5E} \right\rangle$ to the state $\left| {{}^3T} \right\rangle$. Therefore, the orbital ordering with increasing pressure should disappear, and the FM nature of the superexchange will manifest itself (see Tab. 1).

 d$^6$. At the ambient pressure, in the wustite Mg$_x$Fe$_{1 - x}$O  with  cations Fe$^{2 + }$ in the   configuration d$^6$ there is a competition of two different contributions  $J_{{}^4T,{}^4T}^{AFM}$   with   $\sigma$ overlapping and $J_{{}^6A,{}^4T}^{FM}$ with  $\pi$ overlapping, and the AFM contribution dominates. At high pressures (P = 55 GPa), the magnetic moment in the wustite is absent as well as in all other compounds with cations in   configuration d$^6$. The large class of such materials with  $S_{n_0} = 0$ in the ground state is given by the perovskite based rare earth cobaltite LaCoO$_3$, where La$^{3 + }$  is the 4f ion.

d$^8$. For nickel monoxide NiO  with   cations Ni$^{2+}$ in the configuration d$^8$ situation is similar to the  configuration d$^3$. There is no spin crossover in the neutral   configuration d$^8$ and  at the ambient pressure the AFM interaction $J_{{}^2E,{}^2E}^{AFM}$ involves the excited  state $\left| h \right\rangle  = \left| {{}^2E} \right\rangle$ in the hole  configuration d$^7$. Above the spin crossover in the hole configuration this state becomes the ground  one $\left| {{h_0}} \right\rangle  = \left| {{}^2E} \right\rangle$, and the same AFM interaction $J_{{}^2E,{}^2E}^{AFM}$ is given now by the main exchange loop. Thus, its value increases due to the spin crossover in the hole   configuration d$^7$. Summarizing our analysis we get together all our conclusions in Tab.1, and also compare them with experimental data, where it is possible.

\section{\label{sec:VI} Conclusions  \\}
The sign of the partial  contributions ${J_{he}}$  to the total superexchange interaction is directly independent on the cation spin $S\left( {{d^n}} \right)$, but is controlled by the spin relation  $S\left( {{d^{n - 1}}} \right) = S\left( {{d^{n + 1}}} \right)$ (AFM interaction) or   $S\left( {{d^{n - 1}}} \right) = S\left( {{d^{n + 1}}} \right) \pm 1$ (FM interaction) provided that  $S\left( {{d^n}} \right) = S\left( {{d^{n \pm 1}}} \right) \pm {1 \mathord{\left/
 {\vphantom {1 2}} \right.
 \kern-\nulldelimiterspace} 2}$ (see Eqs.(\ref{eq:A.12}) and (\ref{eq:A.14})). Indeed the chromium dioxide CrO$_2$ and nickel monoxide NiO ($S_{n_0}=1$), or manganites LaMnO$_3$ and wustite Mg$_{1-x}$Fe$_x$O ($S_{n_0}=2$), can have FM and AFM interactions respectively. The main factor for the comparison between the AFM and FM interactions is the type  overlapping states involved by the contributions.\\
The nature of the superexchange interaction with increasing pressure changes  (AFM$\rightarrow$FM) only in oxide materials with   cations in d$^5$  (e.g. FeBO$_3$) and d$^7$ (e.g. CoO) configurations. Indeed spin crossover  $\left| {{}^4{T_1}} \right\rangle  \to \left| {{}^2E} \right\rangle$ with generating Jahn-Teller  cations Co$^{2+}\left( {^2E} \right)$ in cobalt monoxide at P $>$ 43 Gpa is accompanied by: (i) transformation of the cubic rock salt-type structure to mixed rhombohedrally distorted rock salt-type structure without significant volume change structure; (ii) a resistance drop by eight orders of magnitude at the room temperature   (43Gpa$<$P$<$63Gpa) while maintaining its semiconductor nature; (iii) Mott-Hubbard transition into the metal rock salt structure at more high pressure P $>$120Gpa.~\cite{Atou2004, Guo2002} We did not find any studies related to high pressure effects in oxides  La$_2$CoO$_4$($S_{n_0}=3/2$,~$T_N=275K$ at the ambient pressure~\cite{Yamada1989}) and LaNiO$_{3-x}$(paramagnetic metal~\cite{Sanchez1996} and ultrathin film AFM insulator~\cite{Golalikhani2018} at the ambient pressure). Unlike the cobalt oxides LaSrCoO$_4$ and LaCoO$_3$, with Co$^{3+}$,~\cite{Lis2019} the layered oxide La$_2$CoO$_4$ has not been studied under the high pressure.
However, these oxide materials~\cite{Drees2014} isostructural to well-known high-T$_C$ and colossal magnetoresistance materials could have interesting physical properties at the high pressure ($>$43Gpa) and magnetic field. On the one hand, the high-T$_C$ superconductors: doped and nonstoichiometric cuprates~\cite{Kampf1994} with the multimode Jahn-Teller  $\left( {{}^2{a_1} + {}^2{b_1}} \right) \otimes \left( {{b_{1g}} + {a_{1g}}} \right)$  effect~\cite{Bersuker1992} and iron based superconductors,~\cite{Si2008} have also the spin ${S_{{n_0}}} = {1 \mathord{\left/
{\vphantom {1 2}} \right.
\kern-\nulldelimiterspace} 2}$ and on the other hand, pseudogap effects and colossal magnetoresistance are observed in the doped manganite La(Sr,Ba)MnO$_3$  also with the Jahn-Teller Mn$^{3+}(^5E)$  cations.~\cite{Saitoh2000}  However, the cobalt oxide La$_2$CoO$_4$ at the high pressure is very likely different from cuprate La$_2$CuO$_4$ at the ambient pressure in a sign of the superexchange interaction, despite the same cation spin $1/2$.
Indeed the interaction in nickel monoxide does not undergo any critical changes with increasing pressure, either in theory or experiment, up to 220 GPa.~\cite{Lyubutin2009} Note, in the  oxide materials: CrO$_2$, NiO, La$_2$CuO$_4$  with the cations in the   electron configurations d$^3$, d$^8$, d$^9$ the spin crossover under high pressure is impossible.

The results partially disagree with experimental data at the ambient pressure only for oxide materials with Jahn-Teller d$^4$  cations like  LaMnO$_3$, where the FM $ab$ planes have AFM ordering. With increasing pressure, the spin crossover in manganite LaMnO$_3$  is accompanied by the transition of the magnetic Jahn-Teller Mn$^{3+}$  cation to the state $\left| {{}^3T} \right\rangle$.  In according to our conclusions, the effects of orbital ordering  should disappear, and the FM nature of the superexchange will manifest itself (see Tab.1). Indeed, below pressure 29 GPa the manganite is not metallic and consists of a dynamic mixture of distorted and undistorted  MnO$_6$ octahedral.~\cite{Baldini2011} Above pressure 32 Gpa, undoped manganite already shows metallic properties.
\begin{acknowledgments}
We acknowledge the support of the Russian Science Foundation through Grant ¹ 18-12-00022.
\end{acknowledgments}

\appendix*
\section{AFM and FM contributions  to superexchange interaction}
To derive Eqs.(\ref{eq:1}) and (\ref{eq:6}), we start from the Hamiltonian of the p-d model, where

$\hat H = {\hat H_d} + {\hat H_p} + {\hat H_{pd}} + {\hat H_{pp}}$
\begin{widetext}
\begin{eqnarray}
{\hat {H_d}} = \sum\limits_{i\lambda \sigma } {\left[ {\left( {{\varepsilon _\lambda } - \mu } \right)d_{\lambda i\sigma }^ + {d_{\lambda i\sigma }} + \frac{{{U_d}}}{2}\hat n_{\lambda i}^\sigma \hat n_{\lambda i}^{ - \sigma } + \frac{1}{2}\sum\limits_{\lambda ' \ne \lambda } {\left( {\sum\limits_{\sigma '} {{V_{\lambda \lambda '}}\hat n_{\lambda i}^\sigma \hat n_{\lambda 'i}^{\sigma '}}  - {J_H}d_{\lambda i\sigma }^ + {d_{\lambda i\bar \sigma }}d_{\lambda 'i\bar \sigma }^ + {d_{\lambda 'i\sigma }}} \right)} } \right]},
\label{eq:A.1} \nonumber \\
{\hat H_p} = \sum\limits_{m\alpha \sigma } {\left[ {\left( {{\varepsilon _\alpha } - \mu } \right)p_{\alpha m\sigma }^ + {p_{\alpha m\sigma }} + \frac{{{U_p}}}{2}\hat n_{\alpha m}^\sigma \hat n_{\alpha m}^{ - \sigma } + \frac{1}{2}\sum\limits_{\alpha ' \ne \alpha ,\sigma '} {{V_{\alpha \alpha '}}\hat n_{\alpha m}^\sigma \hat n_{\alpha 'm}^{\sigma '}} } \right]}, \nonumber \\
{\hat H_{pd}} = \sum\limits_{mi} {\sum\limits_{\alpha \lambda \sigma } {\left[ {t_{im}^{\lambda \alpha }\left( {p_{\alpha m\sigma }^ + {d_{\lambda i\sigma }} + h.c.} \right) + \frac{{V_{im}^{pd}}}{2}\sum\limits_{\sigma '} {\hat n_{\alpha m}^\sigma \hat n_{\lambda i}^{\sigma '}} } \right]} }, ~
{\hat H_{pp}} = \sum\limits_{mn} {\sum\limits_{\alpha \beta \sigma } {t_{mn}^{\alpha \beta }\left( {p_{\alpha m\sigma }^ + {p_{\beta n\sigma }} + h.c.} \right)} }.
\end{eqnarray}
\end{widetext}
Here, $n_{\lambda i}^\sigma  = d_{\lambda i\sigma }^ + {d_{\lambda i\sigma }}$, $n_{\alpha m}^\sigma  = p_{\alpha m\sigma }^ + {p_{\alpha m\sigma }}$, where the indices $i\left( j \right)$  and  $m(n)$ run over all positions ${d_\lambda } = {d_{{x^2} - {y^2}}},{d_{3{r^2} - {z^2}}},{d_{xy}},{d_{xz}},{d_{yz}}$  and  ${p_\alpha } = {p_x},{p_y},{p_z}$ localized one electron states with energies ${\varepsilon _\lambda }$  and ${\varepsilon _\alpha}$; ${t^{\lambda \alpha }_{im}}$   and ${t^{\alpha \beta }_{mn}}$  the hopping matrix elements; ${U_d }$, ${U_p }$    and  ${J_H}$ are one site Coulomb interactions and the Hund exchange interaction, ${V^{pd}_{im}}$  is the energy of repulsion of cation and anion electrons. A correct transition from the Hamiltonian (\ref{eq:A.1}) of the p-d model  to the Hamiltonian (\ref{eq:3}) in the multielectron  representation of the Hubbard operators is possible when constructing well localized Wannier cell oxygen states $\left| {p_{\lambda i\sigma }^ + } \right\rangle$. Although, there is no general derivation of the canonical transformation  $\left| {p_{\lambda i\sigma }^ + } \right\rangle  \leftrightarrow \left| {p_{\alpha m\sigma }^ + } \right\rangle$ for arbitrary lattice symmetry, we assume that the canonical representation does exist and that the Wannier cell oxygen functions are sufficiently localized.~\cite{Shastry1989,Feiner1996,Gavrichkov2006}
In the multielectron representation the one-electron  $p_{\lambda i\sigma }^ +$ and $d_{\lambda i\sigma }^ +$  operators can be written as a superposition of the Hubbard operators that describe one electron excitations from the LS and HS partner states $\left| {h\left( e \right)} \right\rangle$  with spin ${S_{e\left( h \right)}} = {S_n} \pm {1 \mathord{\left/
 {\vphantom {1 2}} \right.
 \kern-\nulldelimiterspace} 2}$  to the neutral state $\left| n \right\rangle$:
 \begin{equation}
 c_{\lambda i\sigma }^ +  = \sum\limits_n {\left[ {\sum\limits_e {\gamma _\lambda ^{}\left( {ne} \right)\alpha _{i\sigma }^ + \left( {ne} \right)}  + \sum\limits_h {\gamma _\lambda ^{}\left( {nh} \right)\beta _{i\sigma }^ + \left( {nh} \right)} } \right]},                                                                                \label{eq:A.2}
 \end{equation}
where the new operators $\alpha _{i\sigma }^ + \left( {en} \right)$  and $\beta _{i\sigma }^ + \left( {nh} \right)$  are notations for the electron addition to the ground state ${N_0} \to {N_ + }$, and to the hole state ${N_ - } \to {N_0}$, respectively. Calculation of the matrix elements in Eq.(\ref{eq:5}) in agreement with the rules of addition of angular momentums results in the following relations:
 \begin{equation}
 \alpha _{i\sigma }^ + \left( {en} \right) = \left\{ {\begin{array}{*{20}{c}}
   {\begin{array}{*{20}{c}}
   {\eta \left( \sigma  \right)\sum\limits_{ - {M_\nu }}^{{M_\nu }} {\sqrt {\frac{{{S_n} - \eta \left( \sigma  \right){M_e} + \tfrac{1}{2}}}{{2{S_n} + 1}}} X_i^{{M_e},{M_n} = {M_e} - \sigma }} } &    \\
\end{array} }  \\
   {\begin{array}{*{20}{c}}
   {\sum\limits_{ - {M_e}}^{{M_e}} {\sqrt {\frac{{{S_n} + \eta \left( \sigma  \right){M_e} + \tfrac{1}{2}}}{{2{S_n} + 1}}} X_i^{{M_e},{M_n} = {M_e} - \sigma }} } &   \\
\end{array} }  \\
\end{array} } \right.
\label{eq:A.3}
\end{equation}
and
\begin{equation}
\beta _{i\sigma }^ + \left( {nh} \right) = \left\{ {\begin{array}{*{20}{c}}
   {\begin{array}{*{20}{c}}
   {\eta \left( \sigma  \right)\sum\limits_{ - {M_n}}^{{M_n}} {\sqrt {\frac{{{S_h} - \eta \left( \sigma  \right){M_n} + \tfrac{1}{2}}}{{2{S_h} + 1}}} X_i^{{M_n},{M_h} = {M_n} - \sigma }} } &  \\
\end{array} }  \\
   {\begin{array}{*{20}{c}}
   {\sum\limits_{ - {M_n}}^{{M_n}} {\sqrt {\frac{{{S_h} + \eta \left( \sigma  \right){M_n} + \tfrac{1}{2}}}{{2{S_h} + 1}}} X_i^{{M_n},{M_h} = {M_n} - \sigma }} } &   \\
\end{array} }  \\
\end{array} } \right.
\label{eq:A.4}
\end{equation}
 where top and below lines are for ${{S_e} = {S_n} - \left| \sigma  \right|}$ and ${{S_e} = {S_n} + \left| \sigma  \right|}$ respectively.
The superexchange interaction appears in the second order of the cell perturbation theory with respect to the hopping processes ${\hat H_1}$  in Eq.(\ref{eq:3}), which corresponds to virtual excitations through the dielectric gap into the conduction band and back to valence band. These quasiparticle excitations correspond to the electron-hole excitations and are described by off-diagonal elements with root vectors  $r = (h,n) $ and $(n,e)$. To highlight these contributions, we use a set of projection operators ${P_h}$  and ${P_e}$ , that generalized the Hubbard model analysis~\cite{Chao1977} to the Mott-Hubbard approach with an arbitrary quasiparticle spectrum, where ${P_h} = \left( {X_i^{hh} + \sum\limits_n {X_i^{nn}} } \right)\left( {X_j^{hh} + \sum\limits_{n'} {X_j^{n'n'}} } \right)$
  and  ${P_e} = X_i^{ee} + X_j^{e'e'} - X_i^{ee}\sum\limits_{e'} {X_j^{e'e'}}$ with $1 \leqslant h \leqslant {N_h}$, $1 \leqslant n \leqslant {N_n}$   and $1 \leqslant e\left( {e'} \right) \leqslant {N_e}$. These operators satisfies the relations $\left( {\sum\limits_{h = 1}^{{N_h}} {{P_h}}  + \sum\limits_{e = 1}^{{N_e}} {{P_e}} } \right) = 1$  and ${P_h}{P_e} = 0$, ${P_h}{P_{h'}} = {\delta _{hh'}}{P_h}$, ${P_e}{P_{e'}} = \delta {}_{ee'}{P_e}$. We introduce the Hamiltonian of the exchange coupled ($i, j$) -pairs: ${\hat h_{ij}} = \left( {{{\hat h}_0} + \hat h_1^{in}} \right) + \hat h_1^{out}$, where  $\left( {{{\hat h}_0} + \hat h_1^{in}} \right) = \sum\limits_{hh'} {{P_h}{{\hat h}_{ij}}{P_{h'}}}  + \sum\limits_{ee'} {{P_e}{{\hat h}_{ij}}} {P_{e'}}$  and  $\hat h_1^{out} = \left( {\sum\limits_h {{P_h}} } \right){\hat h_{ij}}\left( {\sum\limits_e {{P_e}} } \right) + \left( {\sum\limits_e {{P_e}} } \right){\hat h_{ij}}\left( {\sum\limits_h {{P_h}} } \right)$ is the intra- and interband contributions for Hamiltonian   $\hat H_1 = \sum\limits_{ij} {{{\hat h}_{ij}}} $ respectively. In the unitary transformation the Hamiltonian for  ($i,j$) -pairs is equal to ${\tilde h_{ij}} = {e^{\hat G}}{\hat h_{ij}}{e^{ - \hat G}}$ , where $\hat G$  satisfies the equation
 \begin{eqnarray}
&&\left( {\sum\limits_h {{P_h}} } \right){{\hat h}_{ij}}\left( {\sum\limits_e {{P_e}} } \right) + \left( {\sum\limits_e {{P_e}} } \right){{\hat h}_{ij}}\left( {\sum\limits_h {{P_h}} } \right) +
\nonumber \\
&&+ \left[ {\hat G,\left( {\sum\limits_{hh'} {{P_h}{{\hat h}_{ij}}{P_{h'}}}  + \sum\limits_{ee'} {{P_e}{{\hat h}_{ij}}{P_{e'}}} } \right)} \right] = 0,
 \label{eq:A.5}
 \end{eqnarray}
and the transformed Hamiltonian ${\tilde h_{ij}}$  in the second order of cell perturbation theory over interband hopping $\hat h_1^{out}$  can be represented as
 \begin{widetext}
 \begin{equation}
 {\tilde h_{ij}} \approx \left( {\sum\limits_{hh'} {{P_h}} {{\hat h}_{ij}}{P_{h'}} + \sum\limits_{ee'} {{P_e}{{\hat h}_{ij}}{P_{e'}}} } \right) + \frac{1}{2}\left[ {\hat G,\left\{ {\left( {\sum\limits_h {{P_h}} } \right){{\hat h}_{ij}}\left( {\sum\limits_e {{P_e}} } \right) + \left( {\sum\limits_e {{P_e}} } \right){{\hat h}_{ij}}\left( {\sum\limits_h {{P_h}} } \right)} \right\}} \right]
 \label{eq:A.6}
 \end{equation}
 \end{widetext}
 where
 \begin{eqnarray}
&&\left( {\sum\limits_h {{P_h}} } \right){\hat h_{ij}}\left( {\sum\limits_e {{P_e}} } \right) =
\sum\limits_{n\sigma } {\sum\limits_{he} {t_{ij}^{en,hn}\alpha _{i\sigma }^ +
\left( {en} \right)} } \beta _{j\sigma }^{}\left( {hn} \right),
\nonumber \\
&&\left( {\sum\limits_e {{P_e}} } \right){\hat h_{ij}}\left( {\sum\limits_h {{P_h}} } \right) = \sum\limits_{n\sigma } {\sum\limits_{he} {t_{ij}^{ne,nh}\beta _{i\sigma }^ + \left( {nh} \right)\alpha _{j\sigma }^{}\left( {ne} \right)} } \nonumber \\
\label{eq:A.7}
\end{eqnarray}
and
\begin{eqnarray}
\hat G = \sum\limits_{nhe} {\left[ {\frac{{t_{ij}^{en,hn}}}{{{\Delta _{nhe}}}}\sum\limits_\sigma  {\alpha _{i\sigma }^ + \left( {en} \right)\beta _{j\sigma }^{}\left( {hn} \right)}  - } \right.}&&
\nonumber \\
\left. { - \frac{{t_{ij}^{nh,ne}}}{{{\Delta _{{n_0}he}}}}\sum\limits_\sigma  {\beta _{i\sigma }^ + \left( {nh} \right)\alpha _{j\sigma }^{}\left( {ne} \right)} } \right]&&
\label{eq:A.8}
\end{eqnarray}
with the energy denominator is ${\Delta _{nhe}} = \left( {{\varepsilon _e} + {\varepsilon _h}} \right) - 2{\varepsilon _n}$. The effects of the ligand environment of magnetic ions are taken into account, due to the Wannier oxygen cell  functions, as well as the exact diagonalization procedure when constructing the configuration space of the cell $\left| n \right\rangle$  and $\left| {h\left( e \right)} \right\rangle$  states with energies ${\varepsilon _n}$   and ${\varepsilon _{e\left( h \right)}}$  respectively. In agreement with the relations:
\begin{eqnarray}
&&\hat n_{in\sigma }^{\left( e \right)} = \left( {2{S_h} + 1} \right)\beta _{i\sigma }^{\left( t \right) + }\left( {nh} \right)\beta _{i\sigma }^{\left( t \right)}\left( {hn} \right),
\nonumber \\
&&n_{in\sigma }^{\left( h \right)} = \left( {2{S_n} + 1} \right)\alpha _{i\sigma }^{\left( s \right)}\left( {ne} \right)\alpha _{i\sigma }^{\left( s \right) + }\left( {en} \right)
\label{eq:A.9}
\end{eqnarray}

\begin{widetext}
\begin{eqnarray}
&&S_{in}^ +  = \left\{ {\begin{array}{*{20}{c}}
   {\begin{array}{*{20}{c}}
   {\left( {2{S_h} + 1} \right)\beta _{i \uparrow }^ + \left( {nh} \right)\beta _{i \downarrow }^{}\left( {hn} \right) = \left( {2{S_n} + 1} \right)\alpha _ \downarrow ^{}\left( {ne} \right)\alpha _ \downarrow ^ + \left( {en} \right)} & , & {{S_n} = {S_h} + \left| \sigma  \right|;{S_e} = {S_n} + \left| \sigma  \right|}  \\
\end{array} }  \\
   { - \begin{array}{*{20}{c}}
   {\left( {2{S_h} + 1} \right)\beta _{i \uparrow }^ + \left( {nh} \right)\beta _{i \downarrow }^{}\left( {hn} \right) =  - \left( {2{S_n} + 1} \right)\alpha _{i \downarrow }^{}\left( {ne} \right)\alpha _{i \downarrow }^ + \left( {en} \right)} & , & {{S_n} = {S_h} - \left| \sigma  \right|;{S_e} = {S_n} - \left| \sigma  \right|}  \\
\end{array} }  \\
\end{array} } \right. \label{eq:A.10} \\
&&\hat S_{in}^z = \left\{ {\begin{array}{*{20}{c}}
   {\begin{array}{*{20}{c}}
   {\left( {2{S_h} + 1} \right)\sum\limits_\sigma  {\eta \left( \sigma  \right)\beta _{i\sigma }^ + {\beta _{i\sigma }}}  = \left( {2{S_n} + 1} \right)\sum\limits_\sigma  {\eta \left( \sigma  \right){\alpha _{i\sigma }}\alpha _{i\sigma }^ + } } & , & {{S_n} = {S_h} + \left| \sigma  \right|;{S_e} = {S_n} + \left| \sigma  \right|}  \\
\end{array} }  \\
   {\begin{array}{*{20}{c}}
   { - \left( {2{S_h} + 1} \right)\sum\limits_\sigma  {\eta \left( \sigma  \right)\beta _{i\sigma }^ + {\beta _{i\sigma }}}  =  - \left( {2{S_n} + 1} \right)\sum\limits_\sigma  {\eta \left( \sigma  \right){\alpha _{i\sigma }}\alpha _{i\sigma }^ + } } & , & {{S_n} = {S_h} - \left| \sigma  \right|;{S_e} = {S_n} - \left| \sigma  \right|}  \\
\end{array} }  \\
\end{array} } \right. \nonumber
\end{eqnarray}
\end{widetext}
and assuming that the ground state $\left| n \right\rangle  = \left| {{n_0}} \right\rangle$ is occupied at $T = 0K$, and the superexchange Hamiltonian  takes the form:
\begin{equation}
{\hat H_s} = \sum\limits_{i \ne j} {{{\tilde h}_{ij}}}  = \sum\limits_{i \ne j} {\left\{ {J_{ij}^ - {{\hat S}_{i{n_0}}}{{\hat S}_{j{n_0}}} - \frac{1}{4}J_{ij}^ + \hat n_{i{n_0}}^{\left( e \right)}\hat n_{j{n_0}}^{\left( h \right)}} \right\}}
\label{eq:A.11}
\end{equation}

where

\begin{eqnarray}
J_{ij}^ -  = \sum\limits_{he}&& {^{'}{}\frac{{J_{ij}^{}\left( {h,{n_0},e} \right)}}{{\left( {2{S_h} + 1} \right)\left( {2{S_{{n_0}}} + 1} \right)}}} \nonumber \\
&&-  \sum\limits_{he} {{}^{''}\frac{{J_{ij}^{}\left( {h,{n_0},e} \right)}}{{\left( {2{S_h} + 1} \right)\left( {2{S_{{n_0}}} + 1} \right)}}},
\label{eq:A.12}
\end{eqnarray}

and

\begin{eqnarray}
J_{ij}^ +  = \sum\limits_{he}&& {^{'}{}\frac{{J_{ij}^{}\left( {h,{n_0},e} \right)}}{{\left( {2{S_h} + 1} \right)\left( {2{S_{{n_0}}} + 1} \right)}}} + \nonumber \\
&&+ \sum\limits_{he} {{}^{''}\frac{{J_{ij}^{}\left( {h,{n_0},e} \right)}}{{\left( {2{S_h} + 1} \right)\left( {2{S_{{n_0}}} + 1} \right)}}}
\label{eq:A.13}
\end{eqnarray}

and  $\hat n_{i{n_0}}^{\left( e \right)} = \sum\limits_\sigma  {\hat n_{i{n_0}\sigma }^{\left( e \right)}}$, $\hat n_{i{n_0}}^{\left( h \right)} = \sum\limits_\sigma  {\hat n_{i{n_0}\sigma }^{\left( h \right)}}$. Since in the first contribution ($\sum\limits_{he} {{}'...}$) the exchange loops are summed with ${S_h} = {S_e}$, and in the second one ($\sum\limits_{he} {''...} $), the exchange loops are with ${S_h} = {S_e} \pm 1$, so the superexchange ${\hat H_s}$  contains all possible nonzero contributions, and the exchange constant $J^-_{ij}$ in Eq.(\ref{eq:A.11})  is the sum of two AFM and FM contributions.
Note, to obtain Eq.(\ref{eq:A.12}) for the same spins ${S_{i{n_0}}} = {S_{j{n_0}}}$  at the different $i$ and $j$ cell of lattice we could use equality:
\begin{eqnarray}
{J_{ij}} =&& \sum\limits_{he} {{J_{ij}}\left( {h,{n_0},e} \right) = } J_{ij}^{AFM} + J_{ij}^{FM}= \nonumber \\\nonumber \\
=&& 2\sum\limits_{he} {\frac{{{{\left( {t_{ij}^{{n_0}h,{n_0}e}} \right)}^2}\left( {{\delta_{{S_h},{S_e}}} + {\delta _{{S_h},{S_e} \pm 1}}} \right)}}{{{\Delta _{{n_0}he}}}}{\delta _{{S_{{n_0}}},{S_h} + \sigma }}}  \nonumber \\
\label{eq:A.14}
\end{eqnarray}
\\
This is a simple but nonobvious conclusion, since the sign of the exchange parameter $J^{AFM}_{ij}(J^{FM}_{ij})$
becomes clear only after the spin correlators are derived from the operator structure of the Hamiltonian (\ref{eq:2}). Eqs.(\ref{eq:A.12}) and (\ref{eq:A.13}) extends the results of work~\cite{Gavrichkov2017} to the oxide materials with arbitrary transition elements.

\bibliography{referencies_Gavrichkov}

\begin{thebibliography}{63}%
\makeatletter
\providecommand \@ifxundefined [1]{%
 \@ifx{#1\undefined}
}%
\providecommand \@ifnum [1]{%
 \ifnum #1\expandafter \@firstoftwo
 \else \expandafter \@secondoftwo
 \fi
}%
\providecommand \@ifx [1]{%
 \ifx #1\expandafter \@firstoftwo
 \else \expandafter \@secondoftwo
 \fi
}%
\providecommand \natexlab [1]{#1}%
\providecommand \enquote  [1]{``#1''}%
\providecommand \bibnamefont  [1]{#1}%
\providecommand \bibfnamefont [1]{#1}%
\providecommand \citenamefont [1]{#1}%
\providecommand \href@noop [0]{\@secondoftwo}%
\providecommand \href [0]{\begingroup \@sanitize@url \@href}%
\providecommand \@href[1]{\@@startlink{#1}\@@href}%
\providecommand \@@href[1]{\endgroup#1\@@endlink}%
\providecommand \@sanitize@url [0]{\catcode `\\12\catcode `\$12\catcode
  `\&12\catcode `\#12\catcode `\^12\catcode `\_12\catcode `\%12\relax}%
\providecommand \@@startlink[1]{}%
\providecommand \@@endlink[0]{}%
\providecommand \url  [0]{\begingroup\@sanitize@url \@url }%
\providecommand \@url [1]{\endgroup\@href {#1}{\urlprefix }}%
\providecommand \urlprefix  [0]{URL }%
\providecommand \Eprint [0]{\href }%
\providecommand \doibase [0]{http://dx.doi.org/}%
\providecommand \selectlanguage [0]{\@gobble}%
\providecommand \bibinfo  [0]{\@secondoftwo}%
\providecommand \bibfield  [0]{\@secondoftwo}%
\providecommand \translation [1]{[#1]}%
\providecommand \BibitemOpen [0]{}%
\providecommand \bibitemStop [0]{}%
\providecommand \bibitemNoStop [0]{.\EOS\space}%
\providecommand \EOS [0]{\spacefactor3000\relax}%
\providecommand \BibitemShut  [1]{\csname bibitem#1\endcsname}%
\let\auto@bib@innerbib\@empty
\bibitem [{\citenamefont {Anderson}(1959)}]{Anderson1959}%
  \BibitemOpen
  \bibfield  {author} {\bibinfo {author} {\bibfnamefont {P.~W.}\ \bibnamefont
  {Anderson}},\ }\href@noop {} {\bibfield  {journal} {\bibinfo  {journal}
  {Phys. Rev.}\ }\textbf {\bibinfo {volume} {115}},\ \bibinfo {pages} {2}
  (\bibinfo {year} {1959})}\BibitemShut {NoStop}%
\bibitem [{\citenamefont {Tanabe}\ and\ \citenamefont
  {Sugano}(1956)}]{Tanabe1956}%
  \BibitemOpen
  \bibfield  {author} {\bibinfo {author} {\bibfnamefont {Y.}~\bibnamefont
  {Tanabe}}\ and\ \bibinfo {author} {\bibfnamefont {S.}~\bibnamefont
  {Sugano}},\ }\href@noop {} {\bibfield  {journal} {\bibinfo  {journal} {J.
  Phys. Soc. Jpn}\ }\textbf {\bibinfo {volume} {11}},\ \bibinfo {pages} {864}
  (\bibinfo {year} {1956})}\BibitemShut {NoStop}%
\bibitem [{\citenamefont {Mentink}\ and\ \citenamefont
  {Eckstein}(2014)}]{Mentink2014}%
  \BibitemOpen
  \bibfield  {author} {\bibinfo {author} {\bibfnamefont {J.~H.}\ \bibnamefont
  {Mentink}}\ and\ \bibinfo {author} {\bibfnamefont {M.}~\bibnamefont
  {Eckstein}},\ }\href@noop {} {\bibfield  {journal} {\bibinfo  {journal}
  {Phys. Rev. Lett.}\ }\textbf {\bibinfo {volume} {113}},\ \bibinfo {pages}
  {057201} (\bibinfo {year} {2014})}\BibitemShut {NoStop}%
\bibitem [{\citenamefont {Kalashnikova}\ \emph {et~al.}(2015)\citenamefont
  {Kalashnikova}, \citenamefont {Kimel},\ and\ \citenamefont
  {Pisarev}}]{Kalashnikova2015}%
  \BibitemOpen
  \bibfield  {author} {\bibinfo {author} {\bibfnamefont {A.~M.}\ \bibnamefont
  {Kalashnikova}}, \bibinfo {author} {\bibfnamefont {A.~V.}\ \bibnamefont
  {Kimel}}, \ and\ \bibinfo {author} {\bibfnamefont {R.~V.}\ \bibnamefont
  {Pisarev}},\ }\href@noop {} {\bibfield  {journal} {\bibinfo  {journal} {Phys.
  Usp.}\ }\textbf {\bibinfo {volume} {58}},\ \bibinfo {pages} {969–980}
  (\bibinfo {year} {2015})}\BibitemShut {NoStop}%
\bibitem [{\citenamefont {Gavrichkov}\ \emph {et~al.}(2017)\citenamefont
  {Gavrichkov}, \citenamefont {Polukeev},\ and\ \citenamefont
  {Ovchinnikov}}]{Gavrichkov2017}%
  \BibitemOpen
  \bibfield  {author} {\bibinfo {author} {\bibfnamefont {V.~A.}\ \bibnamefont
  {Gavrichkov}}, \bibinfo {author} {\bibfnamefont {S.~I.}\ \bibnamefont
  {Polukeev}}, \ and\ \bibinfo {author} {\bibfnamefont {S.~G.}\ \bibnamefont
  {Ovchinnikov}},\ }\href@noop {} {\bibfield  {journal} {\bibinfo  {journal}
  {Phys. Rev. B}\ }\textbf {\bibinfo {volume} {95}},\ \bibinfo {pages} {144424}
  (\bibinfo {year} {2017})}\BibitemShut {NoStop}%
\bibitem [{\citenamefont {Lamonova}\ \emph {et~al.}(2011)\citenamefont
  {Lamonova}, \citenamefont {Zhitlukhina}, \citenamefont {Babkin},
  \citenamefont {Orel}, \citenamefont {Ovchinnikov},\ and\ \citenamefont
  {Pashkevich}}]{Lamonova2011}%
  \BibitemOpen
  \bibfield  {author} {\bibinfo {author} {\bibfnamefont {K.~V.}\ \bibnamefont
  {Lamonova}}, \bibinfo {author} {\bibfnamefont {E.~S.}\ \bibnamefont
  {Zhitlukhina}}, \bibinfo {author} {\bibfnamefont {R.~Y.}\ \bibnamefont
  {Babkin}}, \bibinfo {author} {\bibfnamefont {S.~M.}\ \bibnamefont {Orel}},
  \bibinfo {author} {\bibfnamefont {S.~G.}\ \bibnamefont {Ovchinnikov}}, \ and\
  \bibinfo {author} {\bibfnamefont {Y.~G.}\ \bibnamefont {Pashkevich}},\
  }\href@noop {} {\bibfield  {journal} {\bibinfo  {journal} {J. Phys. Chem. À}\
  }\textbf {\bibinfo {volume} {115}},\ \bibinfo {pages} {13596–13604} (\bibinfo
  {year} {2011})}\BibitemShut {NoStop}%
\bibitem [{\citenamefont {Lyubutin}\ and\ \citenamefont
  {Gavriliuk}(2009)}]{Lyubutin2009}%
  \BibitemOpen
  \bibfield  {author} {\bibinfo {author} {\bibfnamefont {I.}~\bibnamefont
  {Lyubutin}}\ and\ \bibinfo {author} {\bibfnamefont {A.}~\bibnamefont
  {Gavriliuk}},\ }\href@noop {} {\bibfield  {journal} {\bibinfo  {journal}
  {Phys. Usp.}\ }\textbf {\bibinfo {volume} {52}},\ \bibinfo {pages} {989–1017}
  (\bibinfo {year} {2009})}\BibitemShut {NoStop}%
\bibitem [{\citenamefont {Halder}\ \emph {et~al.}(2002)\citenamefont {Halder},
  \citenamefont {Kepert}, \citenamefont {Moubaraki}, \citenamefont {Murray},\
  and\ \citenamefont {Cashion}}]{Halder2002}%
  \BibitemOpen
  \bibfield  {author} {\bibinfo {author} {\bibfnamefont {G.~J.}\ \bibnamefont
  {Halder}}, \bibinfo {author} {\bibfnamefont {C.~J.}\ \bibnamefont {Kepert}},
  \bibinfo {author} {\bibfnamefont {B.~J.}\ \bibnamefont {Moubaraki}}, \bibinfo
  {author} {\bibfnamefont {K.~S.}\ \bibnamefont {Murray}}, \ and\ \bibinfo
  {author} {\bibfnamefont {J.~D.}\ \bibnamefont {Cashion}},\ }\href@noop {}
  {\bibfield  {journal} {\bibinfo  {journal} {Science}\ }\textbf {\bibinfo
  {volume} {298}},\ \bibinfo {pages} {1762} (\bibinfo {year}
  {2002})}\BibitemShut {NoStop}%
\bibitem [{\citenamefont {Brooker}\ and\ \citenamefont
  {Kitchen}(2009)}]{Brooker2009}%
  \BibitemOpen
  \bibfield  {author} {\bibinfo {author} {\bibfnamefont {S.}~\bibnamefont
  {Brooker}}\ and\ \bibinfo {author} {\bibfnamefont {J.~A.}\ \bibnamefont
  {Kitchen}},\ }\href@noop {} {\bibfield  {journal} {\bibinfo  {journal}
  {Dalton Trans.}\ }\textbf {\bibinfo {volume} {36}},\ \bibinfo {pages} {7331}
  (\bibinfo {year} {2009})}\BibitemShut {NoStop}%
\bibitem [{\citenamefont {Ohkoshi}\ \emph {et~al.}(2011)\citenamefont
  {Ohkoshi}, \citenamefont {Imoto}, \citenamefont {Tsunobuchi}, \citenamefont
  {S.Takano},\ and\ \citenamefont {Tokoro}}]{Ohkoshi2011}%
  \BibitemOpen
  \bibfield  {author} {\bibinfo {author} {\bibfnamefont {S.}~\bibnamefont
  {Ohkoshi}}, \bibinfo {author} {\bibfnamefont {K.}~\bibnamefont {Imoto}},
  \bibinfo {author} {\bibfnamefont {Y.}~\bibnamefont {Tsunobuchi}}, \bibinfo
  {author} {\bibnamefont {S.Takano}}, \ and\ \bibinfo {author} {\bibfnamefont
  {H.}~\bibnamefont {Tokoro}},\ }\href@noop {} {\bibfield  {journal} {\bibinfo
  {journal} {Nat. Chem.}\ }\textbf {\bibinfo {volume} {3}},\ \bibinfo {pages}
  {564} (\bibinfo {year} {2011})}\BibitemShut {NoStop}%
\bibitem [{\citenamefont {Wentzcovitch}\ \emph {et~al.}(2009)\citenamefont
  {Wentzcovitch}, \citenamefont {Justo}, \citenamefont {Wu}, \citenamefont
  {da~Silva}, \citenamefont {Yuen},\ and\ \citenamefont
  {Kohlstedt}}]{Wentzcovitch2009}%
  \BibitemOpen
  \bibfield  {author} {\bibinfo {author} {\bibfnamefont {R.~M.}\ \bibnamefont
  {Wentzcovitch}}, \bibinfo {author} {\bibfnamefont {J.~F.}\ \bibnamefont
  {Justo}}, \bibinfo {author} {\bibfnamefont {Z.}~\bibnamefont {Wu}}, \bibinfo
  {author} {\bibfnamefont {C.~R.~S.}\ \bibnamefont {da~Silva}}, \bibinfo
  {author} {\bibfnamefont {D.~A.}\ \bibnamefont {Yuen}}, \ and\ \bibinfo
  {author} {\bibfnamefont {D.}~\bibnamefont {Kohlstedt}},\ }\href@noop {}
  {\bibfield  {journal} {\bibinfo  {journal} {Proceedings of the National
  Academy of Sciences of the United States of America}\ }\textbf {\bibinfo
  {volume} {106}},\ \bibinfo {pages} {8447} (\bibinfo {year}
  {2009})}\BibitemShut {NoStop}%
\bibitem [{\citenamefont {Hsu}\ \emph {et~al.}(2010)\citenamefont {Hsu},
  \citenamefont {Umemoto}, \citenamefont {Wu},\ and\ \citenamefont
  {Wentzcovitch}}]{Hsu2010}%
  \BibitemOpen
  \bibfield  {author} {\bibinfo {author} {\bibfnamefont {H.}~\bibnamefont
  {Hsu}}, \bibinfo {author} {\bibfnamefont {K.}~\bibnamefont {Umemoto}},
  \bibinfo {author} {\bibfnamefont {Z.}~\bibnamefont {Wu}}, \ and\ \bibinfo
  {author} {\bibfnamefont {R.~M.}\ \bibnamefont {Wentzcovitch}},\ }\href@noop
  {} {\bibfield  {journal} {\bibinfo  {journal} {Reviews in Mineralogy and
  Geochemistry}\ }\textbf {\bibinfo {volume} {71}},\ \bibinfo {pages} {169}
  (\bibinfo {year} {2010})}\BibitemShut {NoStop}%
\bibitem [{\citenamefont {Liu}\ \emph {et~al.}(2014)\citenamefont {Liu},
  \citenamefont {Lin}, \citenamefont {Mao},\ and\ \citenamefont
  {Prakapenka}}]{Liu2014}%
  \BibitemOpen
  \bibfield  {author} {\bibinfo {author} {\bibfnamefont {J.}~\bibnamefont
  {Liu}}, \bibinfo {author} {\bibfnamefont {J.-F.}\ \bibnamefont {Lin}},
  \bibinfo {author} {\bibfnamefont {Z.}~\bibnamefont {Mao}}, \ and\ \bibinfo
  {author} {\bibfnamefont {V.~B.}\ \bibnamefont {Prakapenka}},\ }\href@noop {}
  {\bibfield  {journal} {\bibinfo  {journal} {Am. Mineral.}\ }\textbf {\bibinfo
  {volume} {99}},\ \bibinfo {pages} {84} (\bibinfo {year} {2014})}\BibitemShut
  {NoStop}%
\bibitem [{\citenamefont {Sinmyo}\ \emph {et~al.}(2017)\citenamefont {Sinmyo},
  \citenamefont {McCammon},\ and\ \citenamefont {Dubrovinsky}}]{Sinmyo2017}%
  \BibitemOpen
  \bibfield  {author} {\bibinfo {author} {\bibfnamefont {R.}~\bibnamefont
  {Sinmyo}}, \bibinfo {author} {\bibfnamefont {C.}~\bibnamefont {McCammon}}, \
  and\ \bibinfo {author} {\bibfnamefont {L.}~\bibnamefont {Dubrovinsky}},\
  }\href@noop {} {\bibfield  {journal} {\bibinfo  {journal} {Am. Mineral.}\
  }\textbf {\bibinfo {volume} {102}},\ \bibinfo {pages} {1263–1269} (\bibinfo
  {year} {2017})}\BibitemShut {NoStop}%
\bibitem [{\citenamefont {Marbeuf}\ \emph {et~al.}(2013)\citenamefont
  {Marbeuf}, \citenamefont {Matar}, \citenamefont {Negrier}, \citenamefont
  {Kabalan}, \citenamefont {Letard},\ and\ \citenamefont
  {Guionneau}}]{Marbeuf2013}%
  \BibitemOpen
  \bibfield  {author} {\bibinfo {author} {\bibfnamefont {A.}~\bibnamefont
  {Marbeuf}}, \bibinfo {author} {\bibfnamefont {S.~F.}\ \bibnamefont {Matar}},
  \bibinfo {author} {\bibfnamefont {P.}~\bibnamefont {Negrier}}, \bibinfo
  {author} {\bibfnamefont {L.}~\bibnamefont {Kabalan}}, \bibinfo {author}
  {\bibfnamefont {J.~F.}\ \bibnamefont {Letard}}, \ and\ \bibinfo {author}
  {\bibfnamefont {P.}~\bibnamefont {Guionneau}},\ }\href@noop {} {\bibfield
  {journal} {\bibinfo  {journal} {Chem. Phys.}\ }\textbf {\bibinfo {volume}
  {420}},\ \bibinfo {pages} {25} (\bibinfo {year} {2013})}\BibitemShut
  {NoStop}%
\bibitem [{\citenamefont {Nishino}\ \emph {et~al.}(2007)\citenamefont
  {Nishino}, \citenamefont {Boukheddaden}, \citenamefont {Konishi},\ and\
  \citenamefont {Miyashita}}]{Nishino2007}%
  \BibitemOpen
  \bibfield  {author} {\bibinfo {author} {\bibfnamefont {M.}~\bibnamefont
  {Nishino}}, \bibinfo {author} {\bibfnamefont {K.}~\bibnamefont
  {Boukheddaden}}, \bibinfo {author} {\bibfnamefont {Y.}~\bibnamefont
  {Konishi}}, \ and\ \bibinfo {author} {\bibfnamefont {S.}~\bibnamefont
  {Miyashita}},\ }\href@noop {} {\bibfield  {journal} {\bibinfo  {journal}
  {Phys. Rev. Lett.}\ }\textbf {\bibinfo {volume} {98}},\ \bibinfo {pages}
  {247203} (\bibinfo {year} {2007})}\BibitemShut {NoStop}%
\bibitem [{\citenamefont {Ovchinnikov}\ and\ \citenamefont
  {Zabluda}(2004)}]{Ovchinnikov2004}%
  \BibitemOpen
  \bibfield  {author} {\bibinfo {author} {\bibfnamefont {S.}~\bibnamefont
  {Ovchinnikov}}\ and\ \bibinfo {author} {\bibfnamefont {V.}~\bibnamefont
  {Zabluda}},\ }\href@noop {} {\bibfield  {journal} {\bibinfo  {journal}
  {JETP}\ }\textbf {\bibinfo {volume} {98}},\ \bibinfo {pages} {135} (\bibinfo
  {year} {2004})}\BibitemShut {NoStop}%
\bibitem [{\citenamefont {Saha-Dasgupta}\ and\ \citenamefont
  {Oppeneer}(2014)}]{Saha2014}%
  \BibitemOpen
  \bibfield  {author} {\bibinfo {author} {\bibfnamefont {T.}~\bibnamefont
  {Saha-Dasgupta}}\ and\ \bibinfo {author} {\bibfnamefont {P.~M.}\ \bibnamefont
  {Oppeneer}},\ }\href@noop {} {\bibfield  {journal} {\bibinfo  {journal} {MRS
  Bulletin}\ }\textbf {\bibinfo {volume} {39}},\ \bibinfo {pages} {614}
  (\bibinfo {year} {2014})}\BibitemShut {NoStop}%
\bibitem [{\citenamefont {Gavrichkov}\ \emph {et~al.}(2018)\citenamefont
  {Gavrichkov}, \citenamefont {Polukeev},\ and\ \citenamefont
  {Ovchinnikov}}]{Gavrichkov2018}%
  \BibitemOpen
  \bibfield  {author} {\bibinfo {author} {\bibfnamefont {V.~A.}\ \bibnamefont
  {Gavrichkov}}, \bibinfo {author} {\bibfnamefont {S.~I.}\ \bibnamefont
  {Polukeev}}, \ and\ \bibinfo {author} {\bibfnamefont {S.~G.}\ \bibnamefont
  {Ovchinnikov}},\ }\href@noop {} {\bibfield  {journal} {\bibinfo  {journal}
  {JETP}\ }\textbf {\bibinfo {volume} {127}},\ \bibinfo {pages} {713–720}
  (\bibinfo {year} {2018})}\BibitemShut {NoStop}%
\bibitem [{\citenamefont {Bykova}\ \emph {et~al.}(2016)\citenamefont {Bykova},
  \citenamefont {Dubrovinsky}, \citenamefont {Dubrovinskaia}, \citenamefont
  {Bykov}, \citenamefont {McCammon}, \citenamefont {Ovsyannikov}, \citenamefont
  {Liermann}, \citenamefont {Kupenko}, \citenamefont {Chumakov}, \citenamefont
  {Ruffer}, \citenamefont {Hanfland},\ and\ \citenamefont
  {Prakapenka}}]{Bykova2016}%
  \BibitemOpen
  \bibfield  {author} {\bibinfo {author} {\bibfnamefont {L.}~\bibnamefont
  {Bykova}}, \bibinfo {author} {\bibfnamefont {L.}~\bibnamefont {Dubrovinsky}},
  \bibinfo {author} {\bibfnamefont {N.}~\bibnamefont {Dubrovinskaia}}, \bibinfo
  {author} {\bibfnamefont {M.}~\bibnamefont {Bykov}}, \bibinfo {author}
  {\bibfnamefont {C.}~\bibnamefont {McCammon}}, \bibinfo {author}
  {\bibfnamefont {S.~V.}\ \bibnamefont {Ovsyannikov}}, \bibinfo {author}
  {\bibfnamefont {H.~P.}\ \bibnamefont {Liermann}}, \bibinfo {author}
  {\bibfnamefont {I.}~\bibnamefont {Kupenko}}, \bibinfo {author} {\bibfnamefont
  {A.~I.}\ \bibnamefont {Chumakov}}, \bibinfo {author} {\bibfnamefont
  {R.}~\bibnamefont {Ruffer}}, \bibinfo {author} {\bibfnamefont
  {M.}~\bibnamefont {Hanfland}}, \ and\ \bibinfo {author} {\bibfnamefont
  {V.}~\bibnamefont {Prakapenka}},\ }\href@noop {} {\bibfield  {journal}
  {\bibinfo  {journal} {Nat. Comm.}\ }\textbf {\bibinfo {volume} {7}},\
  \bibinfo {pages} {10661} (\bibinfo {year} {2016})}\BibitemShut {NoStop}%
\bibitem [{\citenamefont {Eskes}\ and\ \citenamefont
  {Jefferson}(1993)}]{Eskes1993}%
  \BibitemOpen
  \bibfield  {author} {\bibinfo {author} {\bibfnamefont {H.}~\bibnamefont
  {Eskes}}\ and\ \bibinfo {author} {\bibfnamefont {J.~H.}\ \bibnamefont
  {Jefferson}},\ }\href@noop {} {\bibfield  {journal} {\bibinfo  {journal}
  {Phys. Rev. B}\ }\textbf {\bibinfo {volume} {48}},\ \bibinfo {pages} {9788}
  (\bibinfo {year} {1993})}\BibitemShut {NoStop}%
\bibitem [{\citenamefont {Ohta}\ \emph {et~al.}(1991)\citenamefont {Ohta},
  \citenamefont {Tohyama},\ and\ \citenamefont {Maekawa}}]{Ohta1991}%
  \BibitemOpen
  \bibfield  {author} {\bibinfo {author} {\bibfnamefont {Y.}~\bibnamefont
  {Ohta}}, \bibinfo {author} {\bibfnamefont {T.}~\bibnamefont {Tohyama}}, \
  and\ \bibinfo {author} {\bibfnamefont {S.}~\bibnamefont {Maekawa}},\
  }\href@noop {} {\bibfield  {journal} {\bibinfo  {journal} {Phys. Rev. Lett.}\
  }\textbf {\bibinfo {volume} {66}},\ \bibinfo {pages} {1228} (\bibinfo {year}
  {1991})}\BibitemShut {NoStop}%
\bibitem [{\citenamefont {Eskes}\ \emph {et~al.}(1989)\citenamefont {Eskes},
  \citenamefont {Sawatzky},\ and\ \citenamefont {Feiner}}]{Eskes1989}%
  \BibitemOpen
  \bibfield  {author} {\bibinfo {author} {\bibfnamefont {H.}~\bibnamefont
  {Eskes}}, \bibinfo {author} {\bibfnamefont {G.~A.}\ \bibnamefont {Sawatzky}},
  \ and\ \bibinfo {author} {\bibfnamefont {L.~F.}\ \bibnamefont {Feiner}},\
  }\href@noop {} {\bibfield  {journal} {\bibinfo  {journal} {Physica C}\
  }\textbf {\bibinfo {volume} {160}},\ \bibinfo {pages} {424} (\bibinfo {year}
  {1989})}\BibitemShut {NoStop}%
\bibitem [{\citenamefont {Stechel}\ and\ \citenamefont
  {Jennison}(1988)}]{Stechel1988}%
  \BibitemOpen
  \bibfield  {author} {\bibinfo {author} {\bibfnamefont {E.~B.}\ \bibnamefont
  {Stechel}}\ and\ \bibinfo {author} {\bibfnamefont {D.~R.}\ \bibnamefont
  {Jennison}},\ }\href@noop {} {\bibfield  {journal} {\bibinfo  {journal}
  {Phys. Rev. B}\ }\textbf {\bibinfo {volume} {38}},\ \bibinfo {pages} {4632}
  (\bibinfo {year} {1988})}\BibitemShut {NoStop}%
\bibitem [{\citenamefont {Bulayevskii}\ \emph {et~al.}(1968)\citenamefont
  {Bulayevskii}, \citenamefont {Nagaev},\ and\ \citenamefont
  {Khomskii}}]{Bulayevskii1968}%
  \BibitemOpen
  \bibfield  {author} {\bibinfo {author} {\bibfnamefont {L.~N.}\ \bibnamefont
  {Bulayevskii}}, \bibinfo {author} {\bibfnamefont {E.~L.}\ \bibnamefont
  {Nagaev}}, \ and\ \bibinfo {author} {\bibfnamefont {D.~I.}\ \bibnamefont
  {Khomskii}},\ }\href@noop {} {\bibfield  {journal} {\bibinfo  {journal}
  {JETP}\ }\textbf {\bibinfo {volume} {27}},\ \bibinfo {pages} {836} (\bibinfo
  {year} {1968})}\BibitemShut {NoStop}%
\bibitem [{\citenamefont {Chao}\ \emph {et~al.}(1977)\citenamefont {Chao},
  \citenamefont {Spalek},\ and\ \citenamefont {Oles}}]{Chao1977}%
  \BibitemOpen
  \bibfield  {author} {\bibinfo {author} {\bibfnamefont {K.~A.}\ \bibnamefont
  {Chao}}, \bibinfo {author} {\bibfnamefont {J.}~\bibnamefont {Spalek}}, \ and\
  \bibinfo {author} {\bibfnamefont {A.~M.}\ \bibnamefont {Oles}},\ }\href@noop
  {} {\bibfield  {journal} {\bibinfo  {journal} {J. Phys. C: Solid State
  Physics}\ }\textbf {\bibinfo {volume} {10}},\ \bibinfo {pages} {L271}
  (\bibinfo {year} {1977})}\BibitemShut {NoStop}%
\bibitem [{\citenamefont {Goodenough}(1963)}]{Goodenough1963}%
  \BibitemOpen
  \bibfield  {author} {\bibinfo {author} {\bibfnamefont {J.~B.}\ \bibnamefont
  {Goodenough}},\ }\href@noop {} {\emph {\bibinfo {title} {Magnetism and the
  Chemical Bond}}}\ (\bibinfo  {publisher} {Interscience (Wiley), New York},\
  \bibinfo {year} {1963})\BibitemShut {NoStop}%
\bibitem [{\citenamefont {Kanamori}(1963)}]{Êànamîri1963}%
  \BibitemOpen
  \bibfield  {author} {\bibinfo {author} {\bibfnamefont {J.}~\bibnamefont
  {Kanamori}},\ }\href@noop {} {\emph {\bibinfo {title} {Magnetism, Vol. 1}}}\
  (\bibinfo  {publisher} {Academic Press Inc., New York},\ \bibinfo {year}
  {1963})\BibitemShut {NoStop}%
\bibitem [{\citenamefont {Gavrichkov}\ \emph {et~al.}(2000)\citenamefont
  {Gavrichkov}, \citenamefont {Ovchinnikov}, \citenamefont {Borisov},\ and\
  \citenamefont {Goryachev}}]{Gavrichkov2000}%
  \BibitemOpen
  \bibfield  {author} {\bibinfo {author} {\bibfnamefont {V.~A.}\ \bibnamefont
  {Gavrichkov}}, \bibinfo {author} {\bibfnamefont {S.~G.}\ \bibnamefont
  {Ovchinnikov}}, \bibinfo {author} {\bibfnamefont {A.~A.}\ \bibnamefont
  {Borisov}}, \ and\ \bibinfo {author} {\bibfnamefont {E.~G.}\ \bibnamefont
  {Goryachev}},\ }\href@noop {} {\bibfield  {journal} {\bibinfo  {journal}
  {JETP}\ }\textbf {\bibinfo {volume} {91}},\ \bibinfo {pages} {369–383}
  (\bibinfo {year} {2000})}\BibitemShut {NoStop}%
\bibitem [{\citenamefont {Ovchinnikov}\ \emph {et~al.}(2012)\citenamefont
  {Ovchinnikov}, \citenamefont {Gavrichkov}, \citenamefont {Korshunov},\ and\
  \citenamefont {Shneyder}}]{Ovchinnikov2012}%
  \BibitemOpen
  \bibfield  {author} {\bibinfo {author} {\bibfnamefont {S.~G.}\ \bibnamefont
  {Ovchinnikov}}, \bibinfo {author} {\bibfnamefont {V.~A.}\ \bibnamefont
  {Gavrichkov}}, \bibinfo {author} {\bibfnamefont {M.~M.}\ \bibnamefont
  {Korshunov}}, \ and\ \bibinfo {author} {\bibfnamefont {E.~I.}\ \bibnamefont
  {Shneyder}},\ }\enquote {\bibinfo {title} {Springer series in solid-state
  sciences, vol. 171},}\ \ (\bibinfo {year} {2012})\ Chap.\ \bibinfo {chapter}
  {LDA+GTB method for band structure calculations in the strongly correlated
  materials, Theoretical methods for Strongly Correlated Systems}, p.\ \bibinfo
  {pages} {147}\BibitemShut {NoStop}%
\bibitem [{\citenamefont {Hubbard}(1964)}]{Hubbard1964}%
  \BibitemOpen
  \bibfield  {author} {\bibinfo {author} {\bibfnamefont {J.}~\bibnamefont
  {Hubbard}},\ }\href@noop {} {\bibfield  {journal} {\bibinfo  {journal} {Proc.
  Roy. Soc. A.}\ }\textbf {\bibinfo {volume} {277(1369)}},\ \bibinfo {pages}
  {237} (\bibinfo {year} {1964})}\BibitemShut {NoStop}%
\bibitem [{\citenamefont {Ovchinnikov}(1997)}]{Ovchinnikov1997}%
  \BibitemOpen
  \bibfield  {author} {\bibinfo {author} {\bibfnamefont {S.}~\bibnamefont
  {Ovchinnikov}},\ }\href@noop {} {\bibfield  {journal} {\bibinfo  {journal}
  {Physics-Uspechi}\ }\textbf {\bibinfo {volume} {40}},\ \bibinfo {pages} {993}
  (\bibinfo {year} {1997})}\BibitemShut {NoStop}%
\bibitem [{\citenamefont {Zaltsev}(1975)}]{Zaltsev1975}%
  \BibitemOpen
  \bibfield  {author} {\bibinfo {author} {\bibfnamefont {R.~O.}\ \bibnamefont
  {Zaltsev}},\ }\href@noop {} {\bibfield  {journal} {\bibinfo  {journal}
  {JETP}\ }\textbf {\bibinfo {volume} {41}},\ \bibinfo {pages} {100} (\bibinfo
  {year} {1975})}\BibitemShut {NoStop}%
\bibitem [{\citenamefont {lrkhin}\ and\ \citenamefont
  {lrkhin}(1993)}]{lrkhin1993}%
  \BibitemOpen
  \bibfield  {author} {\bibinfo {author} {\bibfnamefont {V.~Y.}\ \bibnamefont
  {lrkhin}}\ and\ \bibinfo {author} {\bibfnamefont {Y.~P.}\ \bibnamefont
  {lrkhin}},\ }\href@noop {} {\bibfield  {journal} {\bibinfo  {journal} {JETP}\
  }\textbf {\bibinfo {volume} {104}},\ \bibinfo {pages} {3868} (\bibinfo {year}
  {1993})}\BibitemShut {NoStop}%
\bibitem [{\citenamefont {Irkhin}\ and\ \citenamefont
  {Irkhin}(1994)}]{Irkhin1994}%
  \BibitemOpen
  \bibfield  {author} {\bibinfo {author} {\bibfnamefont {V.~Y.}\ \bibnamefont
  {Irkhin}}\ and\ \bibinfo {author} {\bibfnamefont {Y.~P.}\ \bibnamefont
  {Irkhin}},\ }\href@noop {} {\bibfield  {journal} {\bibinfo  {journal} {Phys.
  Stat. Sol. (b)}\ }\textbf {\bibinfo {volume} {183}},\ \bibinfo {pages} {9}
  (\bibinfo {year} {1994})}\BibitemShut {NoStop}%
\bibitem [{\citenamefont {Gavrichkov}\ \emph {et~al.}(2016)\citenamefont
  {Gavrichkov}, \citenamefont {Pchelkina}, \citenamefont {Nekrasov},\ and\
  \citenamefont {Ovchinnikov}}]{Gavrichkov2016}%
  \BibitemOpen
  \bibfield  {author} {\bibinfo {author} {\bibfnamefont {V.~A.}\ \bibnamefont
  {Gavrichkov}}, \bibinfo {author} {\bibfnamefont {Z.~V.}\ \bibnamefont
  {Pchelkina}}, \bibinfo {author} {\bibfnamefont {I.~A.}\ \bibnamefont
  {Nekrasov}}, \ and\ \bibinfo {author} {\bibfnamefont {S.~G.}\ \bibnamefont
  {Ovchinnikov}},\ }\href@noop {} {\bibfield  {journal} {\bibinfo  {journal}
  {International Journal of Modern Physics B}\ }\textbf {\bibinfo {volume}
  {30}},\ \bibinfo {pages} {1650180} (\bibinfo {year} {2016})}\BibitemShut
  {NoStop}%
\bibitem [{\citenamefont {Gavrichkov}\ and\ \citenamefont
  {Ovchinnikov}(2008)}]{Gavrichkov2008}%
  \BibitemOpen
  \bibfield  {author} {\bibinfo {author} {\bibfnamefont {V.~A.}\ \bibnamefont
  {Gavrichkov}}\ and\ \bibinfo {author} {\bibfnamefont {S.~G.}\ \bibnamefont
  {Ovchinnikov}},\ }\href@noop {} {\bibfield  {journal} {\bibinfo  {journal}
  {Fiz.Tverd.Tela}\ }\textbf {\bibinfo {volume} {50}},\ \bibinfo {pages} {1037}
  (\bibinfo {year} {2008})}\BibitemShut {NoStop}%
\bibitem [{\citenamefont {Ovchinnikov}(2008)}]{Ovchinnikov2008}%
  \BibitemOpen
  \bibfield  {author} {\bibinfo {author} {\bibfnamefont {S.~G.}\ \bibnamefont
  {Ovchinnikov}},\ }\href@noop {} {\bibfield  {journal} {\bibinfo  {journal}
  {JETP}\ }\textbf {\bibinfo {volume} {107}},\ \bibinfo {pages} {140–146}
  (\bibinfo {year} {2008})}\BibitemShut {NoStop}%
\bibitem [{\citenamefont {Kuznetsov}\ \emph {et~al.}(2006)\citenamefont
  {Kuznetsov}, \citenamefont {Almeida}, \citenamefont {Dubrovinsky},
  \citenamefont {Ahuja}, \citenamefont {Kwon}, \citenamefont {Kantor},\ and\
  \citenamefont {Guignot}}]{Kuznetsov2006}%
  \BibitemOpen
  \bibfield  {author} {\bibinfo {author} {\bibfnamefont {A.~Y.}\ \bibnamefont
  {Kuznetsov}}, \bibinfo {author} {\bibfnamefont {J.}~\bibnamefont {Almeida}},
  \bibinfo {author} {\bibfnamefont {L.}~\bibnamefont {Dubrovinsky}}, \bibinfo
  {author} {\bibfnamefont {R.}~\bibnamefont {Ahuja}}, \bibinfo {author}
  {\bibfnamefont {S.~K.}\ \bibnamefont {Kwon}}, \bibinfo {author}
  {\bibfnamefont {I.}~\bibnamefont {Kantor}}, \ and\ \bibinfo {author}
  {\bibfnamefont {N.}~\bibnamefont {Guignot}},\ }\href@noop {} {\bibfield
  {journal} {\bibinfo  {journal} {Journal of Applied Physics}\ }\textbf
  {\bibinfo {volume} {99(5)}},\ \bibinfo {pages} {053909} (\bibinfo {year}
  {2006})}\BibitemShut {NoStop}%
\bibitem [{\citenamefont {Zhou}\ \emph {et~al.}(2011)\citenamefont {Zhou},
  \citenamefont {Alonso}, \citenamefont {Muonz}, \citenamefont
  {Fernandez-Diaz},\ and\ \citenamefont {Goodenough}}]{Zhou2011}%
  \BibitemOpen
  \bibfield  {author} {\bibinfo {author} {\bibfnamefont {J.-S.}\ \bibnamefont
  {Zhou}}, \bibinfo {author} {\bibfnamefont {J.~A.}\ \bibnamefont {Alonso}},
  \bibinfo {author} {\bibfnamefont {A.}~\bibnamefont {Muonz}}, \bibinfo
  {author} {\bibfnamefont {M.~T.}\ \bibnamefont {Fernandez-Diaz}}, \ and\
  \bibinfo {author} {\bibfnamefont {J.~B.}\ \bibnamefont {Goodenough}},\
  }\href@noop {} {\bibfield  {journal} {\bibinfo  {journal} {Phys. Rev. Lett.}\
  }\textbf {\bibinfo {volume} {106}},\ \bibinfo {pages} {057201} (\bibinfo
  {year} {2011})}\BibitemShut {NoStop}%
\bibitem [{\citenamefont {Zhou}\ and\ \citenamefont
  {Goodenough}(2002)}]{Zhou2002}%
  \BibitemOpen
  \bibfield  {author} {\bibinfo {author} {\bibfnamefont {J.-S.}\ \bibnamefont
  {Zhou}}\ and\ \bibinfo {author} {\bibfnamefont {J.~B.}\ \bibnamefont
  {Goodenough}},\ }\href@noop {} {\bibfield  {journal} {\bibinfo  {journal}
  {Phys. Rev. Lett.}\ }\textbf {\bibinfo {volume} {89}},\ \bibinfo {pages}
  {087201} (\bibinfo {year} {2002})}\BibitemShut {NoStop}%
\bibitem [{\citenamefont {Gavriliuk}\ \emph {et~al.}(2004)\citenamefont
  {Gavriliuk}, \citenamefont {Trojan}, \citenamefont {Ovchinnikov},
  \citenamefont {Lyubutin},\ and\ \citenamefont {Sarkisyan}}]{Gavriliuk2004}%
  \BibitemOpen
  \bibfield  {author} {\bibinfo {author} {\bibfnamefont {A.~G.}\ \bibnamefont
  {Gavriliuk}}, \bibinfo {author} {\bibfnamefont {I.~A.}\ \bibnamefont
  {Trojan}}, \bibinfo {author} {\bibfnamefont {S.~G.}\ \bibnamefont
  {Ovchinnikov}}, \bibinfo {author} {\bibfnamefont {I.~S.}\ \bibnamefont
  {Lyubutin}}, \ and\ \bibinfo {author} {\bibfnamefont {V.~A.}\ \bibnamefont
  {Sarkisyan}},\ }\href@noop {} {\bibfield  {journal} {\bibinfo  {journal}
  {JETP}\ }\textbf {\bibinfo {volume} {99}},\ \bibinfo {pages} {566–573}
  (\bibinfo {year} {2004})}\BibitemShut {NoStop}%
\bibitem [{\citenamefont {Lyubutin}\ and\ \citenamefont
  {Ovchinnikov}(2012)}]{Lyubutin2012}%
  \BibitemOpen
  \bibfield  {author} {\bibinfo {author} {\bibfnamefont {I.~S.}\ \bibnamefont
  {Lyubutin}}\ and\ \bibinfo {author} {\bibfnamefont {S.~G.}\ \bibnamefont
  {Ovchinnikov}},\ }\href@noop {} {\bibfield  {journal} {\bibinfo  {journal}
  {JMMM}\ }\textbf {\bibinfo {volume} {324}},\ \bibinfo {pages} {3538–3541}
  (\bibinfo {year} {2012})}\BibitemShut {NoStop}%
\bibitem [{\citenamefont {Lyubutin}\ \emph {et~al.}(2013)\citenamefont
  {Lyubutin}, \citenamefont {Struzhkin}, \citenamefont {Mironovich},
  \citenamefont {Gavriliuk}, \citenamefont {Naumov}, \citenamefont {Lin},
  \citenamefont {Ovchinnikov}, \citenamefont {Sinogeikin}, \citenamefont
  {Chow}, \citenamefont {Xiao},\ and\ \citenamefont {Hemley}}]{Lyubutin2013}%
  \BibitemOpen
  \bibfield  {author} {\bibinfo {author} {\bibfnamefont {I.~S.}\ \bibnamefont
  {Lyubutin}}, \bibinfo {author} {\bibfnamefont {V.~V.}\ \bibnamefont
  {Struzhkin}}, \bibinfo {author} {\bibfnamefont {A.~A.}\ \bibnamefont
  {Mironovich}}, \bibinfo {author} {\bibfnamefont {A.~G.}\ \bibnamefont
  {Gavriliuk}}, \bibinfo {author} {\bibfnamefont {P.~G.}\ \bibnamefont
  {Naumov}}, \bibinfo {author} {\bibfnamefont {J.~F.}\ \bibnamefont {Lin}},
  \bibinfo {author} {\bibfnamefont {S.~G.}\ \bibnamefont {Ovchinnikov}},
  \bibinfo {author} {\bibfnamefont {S.}~\bibnamefont {Sinogeikin}}, \bibinfo
  {author} {\bibfnamefont {P.}~\bibnamefont {Chow}}, \bibinfo {author}
  {\bibfnamefont {Y.}~\bibnamefont {Xiao}}, \ and\ \bibinfo {author}
  {\bibfnamefont {R.~J.}\ \bibnamefont {Hemley}},\ }\href@noop {} {\bibfield
  {journal} {\bibinfo  {journal} {Proceedings of National Academy of Science of
  USA (PNAS)}\ }\textbf {\bibinfo {volume} {110(18)}},\ \bibinfo {pages} {7142}
  (\bibinfo {year} {2013})}\BibitemShut {NoStop}%
\bibitem [{\citenamefont {Jauch}\ \emph {et~al.}(2001)\citenamefont {Jauch},
  \citenamefont {Reehuis}, \citenamefont {Bleif}, \citenamefont {Kubanek},\
  and\ \citenamefont {Pattison}}]{Jauch2001}%
  \BibitemOpen
  \bibfield  {author} {\bibinfo {author} {\bibfnamefont {W.}~\bibnamefont
  {Jauch}}, \bibinfo {author} {\bibfnamefont {M.}~\bibnamefont {Reehuis}},
  \bibinfo {author} {\bibfnamefont {H.~J.}\ \bibnamefont {Bleif}}, \bibinfo
  {author} {\bibfnamefont {F.}~\bibnamefont {Kubanek}}, \ and\ \bibinfo
  {author} {\bibfnamefont {P.}~\bibnamefont {Pattison}},\ }\href@noop {}
  {\bibfield  {journal} {\bibinfo  {journal} {Phys. Rev. B}\ }\textbf {\bibinfo
  {volume} {64}},\ \bibinfo {pages} {052102} (\bibinfo {year}
  {2001})}\BibitemShut {NoStop}%
\bibitem [{\citenamefont {Atou}\ \emph {et~al.}(2004)\citenamefont {Atou},
  \citenamefont {Kawasaki},\ and\ \citenamefont {Nakajima}}]{Atou2004}%
  \BibitemOpen
  \bibfield  {author} {\bibinfo {author} {\bibfnamefont {T.}~\bibnamefont
  {Atou}}, \bibinfo {author} {\bibfnamefont {M.}~\bibnamefont {Kawasaki}}, \
  and\ \bibinfo {author} {\bibfnamefont {S.}~\bibnamefont {Nakajima}},\
  }\href@noop {} {\bibfield  {journal} {\bibinfo  {journal} {Jpn. J. Appl.
  Phys.}\ }\textbf {\bibinfo {volume} {43}},\ \bibinfo {pages} {L1281–L1283}
  (\bibinfo {year} {2004})}\BibitemShut {NoStop}%
\bibitem [{\citenamefont {Guo}\ \emph {et~al.}(2002)\citenamefont {Guo},
  \citenamefont {Mao}, \citenamefont {Hu}, \citenamefont {Shu},\ and\
  \citenamefont {Hemley}}]{Guo2002}%
  \BibitemOpen
  \bibfield  {author} {\bibinfo {author} {\bibfnamefont {Q.}~\bibnamefont
  {Guo}}, \bibinfo {author} {\bibfnamefont {H.-K.}\ \bibnamefont {Mao}},
  \bibinfo {author} {\bibfnamefont {J.}~\bibnamefont {Hu}}, \bibinfo {author}
  {\bibfnamefont {J.}~\bibnamefont {Shu}}, \ and\ \bibinfo {author}
  {\bibfnamefont {R.~J.}\ \bibnamefont {Hemley}},\ }\href@noop {} {\bibfield
  {journal} {\bibinfo  {journal} {J. Phys.: Condens. Matter}\ }\textbf
  {\bibinfo {volume} {14}},\ \bibinfo {pages} {11369 – 11374} (\bibinfo {year}
  {2002})}\BibitemShut {NoStop}%
\bibitem [{\citenamefont {Rinaldi-Montes}\ \emph {et~al.}(2016)\citenamefont
  {Rinaldi-Montes}, \citenamefont {Gorria}, \citenamefont {Martinez-Blanco},
  \citenamefont {Fuertes}, \citenamefont {Puente-Orench}, \citenamefont
  {Olivi},\ and\ \citenamefont {Blanco}}]{Rinaldi-Montes2016}%
  \BibitemOpen
  \bibfield  {author} {\bibinfo {author} {\bibfnamefont {N.}~\bibnamefont
  {Rinaldi-Montes}}, \bibinfo {author} {\bibfnamefont {P.}~\bibnamefont
  {Gorria}}, \bibinfo {author} {\bibfnamefont {D.}~\bibnamefont
  {Martinez-Blanco}}, \bibinfo {author} {\bibfnamefont {A.~B.}\ \bibnamefont
  {Fuertes}}, \bibinfo {author} {\bibnamefont {Puente-Orench}}, \bibinfo
  {author} {\bibfnamefont {L.}~\bibnamefont {Olivi}}, \ and\ \bibinfo {author}
  {\bibfnamefont {J.~A.}\ \bibnamefont {Blanco}},\ }\href@noop {} {\bibfield
  {journal} {\bibinfo  {journal} {AIP Advances}\ }\textbf {\bibinfo {volume}
  {6}},\ \bibinfo {pages} {056104} (\bibinfo {year} {2016})}\BibitemShut
  {NoStop}%
\bibitem [{\citenamefont {Kugel}\ and\ \citenamefont
  {Khomskii}(1982)}]{Kugel1982}%
  \BibitemOpen
  \bibfield  {author} {\bibinfo {author} {\bibfnamefont {K.~I.}\ \bibnamefont
  {Kugel}}\ and\ \bibinfo {author} {\bibfnamefont {D.~I.}\ \bibnamefont
  {Khomskii}},\ }\href@noop {} {\bibfield  {journal} {\bibinfo  {journal} {Sov.
  Phys. Usp.}\ }\textbf {\bibinfo {volume} {25(4)}},\ \bibinfo {pages} {231}
  (\bibinfo {year} {1982})}\BibitemShut {NoStop}%
\bibitem [{\citenamefont {Streltsov}\ and\ \citenamefont
  {Khomskii}(2017)}]{Streltsov2017}%
  \BibitemOpen
  \bibfield  {author} {\bibinfo {author} {\bibfnamefont {S.~V.}\ \bibnamefont
  {Streltsov}}\ and\ \bibinfo {author} {\bibfnamefont {D.~I.}\ \bibnamefont
  {Khomskii}},\ }\href@noop {} {\bibfield  {journal} {\bibinfo  {journal}
  {Phys. Usp.}\ }\textbf {\bibinfo {volume} {60}},\ \bibinfo {pages} {1121}
  (\bibinfo {year} {2017})}\BibitemShut {NoStop}%
\bibitem [{\citenamefont {Yamada}\ \emph {et~al.}(1989)\citenamefont {Yamada},
  \citenamefont {Matsuda}, \citenamefont {Endoh}, \citenamefont {Keimer},
  \citenamefont {Birgeneau}, \citenamefont {Onodera}, \citenamefont {Mizusaki},
  \citenamefont {Matsuura},\ and\ \citenamefont {Shirane}}]{Yamada1989}%
  \BibitemOpen
  \bibfield  {author} {\bibinfo {author} {\bibfnamefont {K.}~\bibnamefont
  {Yamada}}, \bibinfo {author} {\bibfnamefont {M.}~\bibnamefont {Matsuda}},
  \bibinfo {author} {\bibfnamefont {Y.}~\bibnamefont {Endoh}}, \bibinfo
  {author} {\bibfnamefont {B.}~\bibnamefont {Keimer}}, \bibinfo {author}
  {\bibfnamefont {R.~J.}\ \bibnamefont {Birgeneau}}, \bibinfo {author}
  {\bibfnamefont {S.}~\bibnamefont {Onodera}}, \bibinfo {author} {\bibfnamefont
  {J.}~\bibnamefont {Mizusaki}}, \bibinfo {author} {\bibfnamefont
  {T.}~\bibnamefont {Matsuura}}, \ and\ \bibinfo {author} {\bibfnamefont
  {G.}~\bibnamefont {Shirane}},\ }\href@noop {} {\bibfield  {journal} {\bibinfo
   {journal} {Phys. Rev. B}\ }\textbf {\bibinfo {volume} {39}},\ \bibinfo
  {pages} {2336} (\bibinfo {year} {1989})}\BibitemShut {NoStop}%
\bibitem [{\citenamefont {Sanchez}\ \emph {et~al.}(1996)\citenamefont
  {Sanchez}, \citenamefont {Causa}, \citenamefont {Caneiro}, \citenamefont
  {Butera}, \citenamefont {Vallet-Regi}, \citenamefont {Sayagues},
  \citenamefont {Gonzalez-Calbet}, \citenamefont {Garcia-Sanz},\ and\
  \citenamefont {Rivas}}]{Sanchez1996}%
  \BibitemOpen
  \bibfield  {author} {\bibinfo {author} {\bibfnamefont {R.~D.}\ \bibnamefont
  {Sanchez}}, \bibinfo {author} {\bibfnamefont {M.~T.}\ \bibnamefont {Causa}},
  \bibinfo {author} {\bibfnamefont {A.}~\bibnamefont {Caneiro}}, \bibinfo
  {author} {\bibfnamefont {A.}~\bibnamefont {Butera}}, \bibinfo {author}
  {\bibfnamefont {M.}~\bibnamefont {Vallet-Regi}}, \bibinfo {author}
  {\bibfnamefont {M.~J.}\ \bibnamefont {Sayagues}}, \bibinfo {author}
  {\bibfnamefont {J.}~\bibnamefont {Gonzalez-Calbet}}, \bibinfo {author}
  {\bibfnamefont {F.}~\bibnamefont {Garcia-Sanz}}, \ and\ \bibinfo {author}
  {\bibfnamefont {J.}~\bibnamefont {Rivas}},\ }\href@noop {} {\bibfield
  {journal} {\bibinfo  {journal} {Phys. Rev. B}\ }\textbf {\bibinfo {volume}
  {54}},\ \bibinfo {pages} {16574} (\bibinfo {year} {1996})}\BibitemShut
  {NoStop}%
\bibitem [{\citenamefont {Golalikhani}\ \emph {et~al.}(2018)\citenamefont
  {Golalikhani}, \citenamefont {Lei}, \citenamefont {Chandrasena},
  \citenamefont {Kasaei}, \citenamefont {Park}, \citenamefont {Bai},
  \citenamefont {Orgiani}, \citenamefont {J.~Ciston}, \citenamefont {Arena},
  \citenamefont {Shafer}, \citenamefont {Arenholz}, \citenamefont {Davidson},
  \citenamefont {Millis}, \citenamefont {Gray},\ and\ \citenamefont
  {Xi}}]{Golalikhani2018}%
  \BibitemOpen
  \bibfield  {author} {\bibinfo {author} {\bibfnamefont {M.}~\bibnamefont
  {Golalikhani}}, \bibinfo {author} {\bibfnamefont {Q.}~\bibnamefont {Lei}},
  \bibinfo {author} {\bibfnamefont {R.~U.}\ \bibnamefont {Chandrasena}},
  \bibinfo {author} {\bibfnamefont {L.}~\bibnamefont {Kasaei}}, \bibinfo
  {author} {\bibfnamefont {H.}~\bibnamefont {Park}}, \bibinfo {author}
  {\bibfnamefont {J.}~\bibnamefont {Bai}}, \bibinfo {author} {\bibfnamefont
  {P.}~\bibnamefont {Orgiani}}, \bibinfo {author} {\bibfnamefont {G.~E.~S.}\
  \bibnamefont {J.~Ciston}}, \bibinfo {author} {\bibfnamefont {D.~A.}\
  \bibnamefont {Arena}}, \bibinfo {author} {\bibfnamefont {P.}~\bibnamefont
  {Shafer}}, \bibinfo {author} {\bibfnamefont {E.}~\bibnamefont {Arenholz}},
  \bibinfo {author} {\bibfnamefont {B.~A.}\ \bibnamefont {Davidson}}, \bibinfo
  {author} {\bibfnamefont {A.~J.}\ \bibnamefont {Millis}}, \bibinfo {author}
  {\bibfnamefont {A.~X.}\ \bibnamefont {Gray}}, \ and\ \bibinfo {author}
  {\bibfnamefont {X.~X.}\ \bibnamefont {Xi}},\ }\href@noop {} {\bibfield
  {journal} {\bibinfo  {journal} {Nat. Comm.}\ }\textbf {\bibinfo {volume}
  {9}},\ \bibinfo {pages} {2206} (\bibinfo {year} {2018})}\BibitemShut
  {NoStop}%
\bibitem [{\citenamefont {Lis}\ \emph {et~al.}(2019)\citenamefont {Lis},
  \citenamefont {E.Kichanov}, \citenamefont {P.Kozlenko}, \citenamefont
  {Jirak}, \citenamefont {V.Belushkin},\ and\ \citenamefont
  {.N.Savenko}}]{Lis2019}%
  \BibitemOpen
  \bibfield  {author} {\bibinfo {author} {\bibfnamefont {O.~N.}\ \bibnamefont
  {Lis}}, \bibinfo {author} {\bibfnamefont {S.}~\bibnamefont {E.Kichanov}},
  \bibinfo {author} {\bibfnamefont {D.}~\bibnamefont {P.Kozlenko}}, \bibinfo
  {author} {\bibfnamefont {Z.}~\bibnamefont {Jirak}}, \bibinfo {author}
  {\bibfnamefont {A.}~\bibnamefont {V.Belushkin}}, \ and\ \bibinfo {author}
  {\bibfnamefont {B.}~\bibnamefont {.N.Savenko}},\ }\href@noop {} {\bibfield
  {journal} {\bibinfo  {journal} {JMMM}\ }\textbf {\bibinfo {volume} {487}},\
  \bibinfo {pages} {165360} (\bibinfo {year} {2019})}\BibitemShut {NoStop}%
\bibitem [{\citenamefont {Drees}\ \emph {et~al.}(2014)\citenamefont {Drees},
  \citenamefont {Li}, \citenamefont {Ricci}, \citenamefont {Rotter},
  \citenamefont {Schmidt}, \citenamefont {Lamago}, \citenamefont {Sobolev},
  \citenamefont {Rutt}, \citenamefont {Gutowski}, \citenamefont {Sprung},
  \citenamefont {Piovano}, \citenamefont {Castellan},\ and\ \citenamefont
  {Komarek}}]{Drees2014}%
  \BibitemOpen
  \bibfield  {author} {\bibinfo {author} {\bibfnamefont {Y.}~\bibnamefont
  {Drees}}, \bibinfo {author} {\bibfnamefont {Z.~W.}\ \bibnamefont {Li}},
  \bibinfo {author} {\bibfnamefont {A.}~\bibnamefont {Ricci}}, \bibinfo
  {author} {\bibfnamefont {M.}~\bibnamefont {Rotter}}, \bibinfo {author}
  {\bibfnamefont {W.}~\bibnamefont {Schmidt}}, \bibinfo {author} {\bibfnamefont
  {D.}~\bibnamefont {Lamago}}, \bibinfo {author} {\bibfnamefont
  {O.}~\bibnamefont {Sobolev}}, \bibinfo {author} {\bibfnamefont
  {U.}~\bibnamefont {Rutt}}, \bibinfo {author} {\bibfnamefont {O.}~\bibnamefont
  {Gutowski}}, \bibinfo {author} {\bibfnamefont {M.}~\bibnamefont {Sprung}},
  \bibinfo {author} {\bibfnamefont {A.}~\bibnamefont {Piovano}}, \bibinfo
  {author} {\bibfnamefont {J.~P.}\ \bibnamefont {Castellan}}, \ and\ \bibinfo
  {author} {\bibfnamefont {A.~C.}\ \bibnamefont {Komarek}},\ }\href@noop {}
  {\bibfield  {journal} {\bibinfo  {journal} {Nat. Comm.}\ }\textbf {\bibinfo
  {volume} {5}},\ \bibinfo {pages} {5731} (\bibinfo {year} {2014})}\BibitemShut
  {NoStop}%
\bibitem [{\citenamefont {Kampf}(1994)}]{Kampf1994}%
  \BibitemOpen
  \bibfield  {author} {\bibinfo {author} {\bibfnamefont {A.~P.}\ \bibnamefont
  {Kampf}},\ }\href@noop {} {\bibfield  {journal} {\bibinfo  {journal} {Physics
  Reports}\ }\textbf {\bibinfo {volume} {249}},\ \bibinfo {pages} {219—351}
  (\bibinfo {year} {1994})}\BibitemShut {NoStop}%
\bibitem [{\citenamefont {Bersuker}\ \emph {et~al.}(1992)\citenamefont
  {Bersuker}, \citenamefont {Gorinchoy}, \citenamefont {Polinger},\ and\
  \citenamefont {Solonenko}}]{Bersuker1992}%
  \BibitemOpen
  \bibfield  {author} {\bibinfo {author} {\bibfnamefont {G.~I.}\ \bibnamefont
  {Bersuker}}, \bibinfo {author} {\bibfnamefont {N.~N.}\ \bibnamefont
  {Gorinchoy}}, \bibinfo {author} {\bibfnamefont {V.~Z.}\ \bibnamefont
  {Polinger}}, \ and\ \bibinfo {author} {\bibfnamefont {A.~O.}\ \bibnamefont
  {Solonenko}},\ }\href@noop {} {\bibfield  {journal} {\bibinfo  {journal}
  {Supercond.: Phys., Chem., Eng.}\ }\textbf {\bibinfo {volume} {5}},\ \bibinfo
  {pages} {1003} (\bibinfo {year} {1992})}\BibitemShut {NoStop}%
\bibitem [{\citenamefont {Si}\ and\ \citenamefont {Abrahams}(2008)}]{Si2008}%
  \BibitemOpen
  \bibfield  {author} {\bibinfo {author} {\bibfnamefont {Q.}~\bibnamefont
  {Si}}\ and\ \bibinfo {author} {\bibfnamefont {E.}~\bibnamefont {Abrahams}},\
  }\href@noop {} {\bibfield  {journal} {\bibinfo  {journal} {PRL}\ }\textbf
  {\bibinfo {volume} {101}},\ \bibinfo {pages} {076401} (\bibinfo {year}
  {2008})}\BibitemShut {NoStop}%
\bibitem [{\citenamefont {Saitoh}\ \emph {et~al.}(2000)\citenamefont {Saitoh},
  \citenamefont {Dessau}, \citenamefont {Moritomo}, \citenamefont {Kimura},
  \citenamefont {Tokura},\ and\ \citenamefont {Hamada}}]{Saitoh2000}%
  \BibitemOpen
  \bibfield  {author} {\bibinfo {author} {\bibfnamefont {T.}~\bibnamefont
  {Saitoh}}, \bibinfo {author} {\bibfnamefont {D.~S.}\ \bibnamefont {Dessau}},
  \bibinfo {author} {\bibfnamefont {Y.}~\bibnamefont {Moritomo}}, \bibinfo
  {author} {\bibfnamefont {T.}~\bibnamefont {Kimura}}, \bibinfo {author}
  {\bibfnamefont {Y.}~\bibnamefont {Tokura}}, \ and\ \bibinfo {author}
  {\bibfnamefont {N.}~\bibnamefont {Hamada}},\ }\href@noop {} {\bibfield
  {journal} {\bibinfo  {journal} {Phys. Rev. B}\ }\textbf {\bibinfo {volume}
  {62}},\ \bibinfo {pages} {1039} (\bibinfo {year} {2000})}\BibitemShut
  {NoStop}%
\bibitem [{\citenamefont {Baldini}\ \emph {et~al.}(2011)\citenamefont
  {Baldini}, \citenamefont {Struzhkin}, \citenamefont {Goncharov},
  \citenamefont {Postorino},\ and\ \citenamefont {Mao}}]{Baldini2011}%
  \BibitemOpen
  \bibfield  {author} {\bibinfo {author} {\bibfnamefont {M.}~\bibnamefont
  {Baldini}}, \bibinfo {author} {\bibfnamefont {V.~V.}\ \bibnamefont
  {Struzhkin}}, \bibinfo {author} {\bibfnamefont {A.~F.}\ \bibnamefont
  {Goncharov}}, \bibinfo {author} {\bibfnamefont {P.}~\bibnamefont
  {Postorino}}, \ and\ \bibinfo {author} {\bibfnamefont {W.~L.}\ \bibnamefont
  {Mao}},\ }\href@noop {} {\bibfield  {journal} {\bibinfo  {journal} {Phys.
  Rev. Lett.}\ }\textbf {\bibinfo {volume} {106}},\ \bibinfo {pages} {066402}
  (\bibinfo {year} {2011})}\BibitemShut {NoStop}%
\bibitem [{\citenamefont {Shastry}(1989)}]{Shastry1989}%
  \BibitemOpen
  \bibfield  {author} {\bibinfo {author} {\bibfnamefont {B.~S.}\ \bibnamefont
  {Shastry}},\ }\href@noop {} {\bibfield  {journal} {\bibinfo  {journal} {Phys.
  Rev. Lett.}\ }\textbf {\bibinfo {volume} {63}},\ \bibinfo {pages} {1288}
  (\bibinfo {year} {1989})}\BibitemShut {NoStop}%
\bibitem [{\citenamefont {Feiner}\ \emph {et~al.}(1996)\citenamefont {Feiner},
  \citenamefont {Jefferson},\ and\ \citenamefont {Raimondi}}]{Feiner1996}%
  \BibitemOpen
  \bibfield  {author} {\bibinfo {author} {\bibfnamefont {L.~F.}\ \bibnamefont
  {Feiner}}, \bibinfo {author} {\bibfnamefont {J.~H.}\ \bibnamefont
  {Jefferson}}, \ and\ \bibinfo {author} {\bibfnamefont {R.}~\bibnamefont
  {Raimondi}},\ }\href@noop {} {\bibfield  {journal} {\bibinfo  {journal}
  {Phys. Rev. B}\ }\textbf {\bibinfo {volume} {53}},\ \bibinfo {pages} {8751}
  (\bibinfo {year} {1996})}\BibitemShut {NoStop}%
\bibitem [{\citenamefont {Gavrichkov}\ \emph {et~al.}(2006)\citenamefont
  {Gavrichkov}, \citenamefont {Ovchinnikov},\ and\ \citenamefont
  {Yakimov}}]{Gavrichkov2006}%
  \BibitemOpen
  \bibfield  {author} {\bibinfo {author} {\bibfnamefont {V.~A.}\ \bibnamefont
  {Gavrichkov}}, \bibinfo {author} {\bibfnamefont {S.~G.}\ \bibnamefont
  {Ovchinnikov}}, \ and\ \bibinfo {author} {\bibfnamefont {L.~E.}\ \bibnamefont
  {Yakimov}},\ }\href@noop {} {\bibfield  {journal} {\bibinfo  {journal}
  {JETP}\ }\textbf {\bibinfo {volume} {102}},\ \bibinfo {pages} {972–985}
  (\bibinfo {year} {2006})}\BibitemShut {NoStop}%
\end{thebibliography}%
\end{document}